\documentclass[12pt,a4paper]{article}
\usepackage[utf8]{inputenc}

\def\papertitle{\bf $\Lambda_b\to p$ transition form factors in perturbative QCD }
\usepackage{authblk}
\title{\papertitle}
\author[1,2]{Jia-Jie Han\footnote{Email: hanjiajie1020@163.com}}
\author[3]{Ya Li\footnote{Email: liyakelly@163.com, corresponding author}}
\author[4]{Hsiang-nan Li\footnote{Email: hnli@phys.sinica.edu.tw, corresponding author}}
\author[5]{Yue-Long Shen\footnote{Email: shenylmeteor@ouc.edu.cn, corresponding author}}
\author[1]{Zhen-Jun Xiao\footnote{Email: xiaozhenjun@njnu.edu.cn, corresponding author}}
\author[2,6,7]{Fu-Sheng Yu\footnote{Email: yufsh@lzu.edu.cn, corresponding author}}
\affil[1]{Department of Physics and Institute of Theoretical Physics,
	Nanjing Normal University, Nanjing 210023, People’s Republic of China}
\affil[2]{School of Nuclear Science and Technology,  and Frontiers Science Center for Rare Isotopes, Lanzhou University, Lanzhou 730000, People’s Republic of China}
\affil[3]{Department of Physics, College of Sciences, Nanjing Agricultural University, Nanjing 210095, People’s Republic of China}
\affil[4]{Institute of Physics, Academia Sinica, Taipei, Taiwan 115, Republic of China}
\affil[5]{College of Physics and Photoelectric Engineering,
	Ocean University of China, Qingdao 266100, People’s Republic of China}
\affil[6]{Lanzhou Center for Theoretical Physics, and Key Laboratory of Theoretical Physics of Gansu Province, Lanzhou University, Lanzhou 730000, People’s Republic of China}
\affil[7]{Center for High Energy Physics, Peking University, Beijing 100871, People’s Republic of China}
\date{\today}                   

\usepackage[top=1in, bottom=1.25in, left=1in, right=1in]{geometry}
\usepackage{microtype}
\usepackage{lineno}
\usepackage{xspace}
\usepackage{booktabs}
\usepackage{comment}

\usepackage{caption} 

\usepackage{graphicx}  
\usepackage{subfigure}
\usepackage{color}
\usepackage{colortbl}
\DeclareGraphicsExtensions{.pdf,.PDF,png,.PNG}

\usepackage{amsmath} 
\usepackage{amssymb}
\usepackage{slashed}
\usepackage{yhmath}
\usepackage{braket}
\usepackage{amsfonts}
\usepackage{amsmath,amsfonts,amssymb,amsthm,bm,upgreek}
\usepackage{upgreek} 

\usepackage{hyperxmp}
\usepackage{hyperref}
\usepackage{hypcap} 
\usepackage[figuresright]{rotating}
\usepackage{setspace}

\def\pion   {{\ensuremath{\pi}}\xspace}

\def\pip    {{\ensuremath{\pion^+}}\xspace}
\def\pim    {{\ensuremath{\pion^-}}\xspace}

\def\kaon   {{\ensuremath{K}}\xspace}
\def\Kbar    {{\kern 0.2em\overline{\kern -0.2em \kaon}{}}\xspace}

\def\KorKbar    {\kern 0.18em\optbar{\kern -0.18em K}{}\xspace}

\def\Km      {{\ensuremath{\kaon^-}}\xspace}

\def\Dbar    {{\kern 0.2em\overline{\kern -0.2em D}{}}\xspace}

\def\DorDbar    {\kern 0.18em\optbar{\kern -0.18em D}{}\xspace}

\def\Lz          {{\ensuremath{\Lambda}}\xspace}
\def\Lbar        {{\ensuremath{\kern 0.1em\overline{\kern -0.1em\Lambda}}}\xspace}
\def\LorLbar    {\kern 0.18em\optbar{\kern -0.18em \PLambda}{}\xspace}

\def\Lb      {{\ensuremath{\Lz^0_b}}\xspace}

\newcommand{\tev}{\ensuremath{\mathrm{\,Te\kern -0.1em V}}\xspace}
\newcommand{\gev}{\ensuremath{\mathrm{\,Ge\kern -0.1em V}}\xspace}
\newcommand{\mev}{\ensuremath{\mathrm{\,Me\kern -0.1em V}}\xspace}
\newcommand{\kev}{\ensuremath{\mathrm{\,ke\kern -0.1em V}}\xspace}
\newcommand{\ev}{\ensuremath{\mathrm{\,e\kern -0.1em V}}\xspace}
\newcommand{\gevc}{\ensuremath{{\mathrm{\,Ge\kern -0.1em V\!/}c}}\xspace}
\newcommand{\mevc}{\ensuremath{{\mathrm{\,Me\kern -0.1em V\!/}c}}\xspace}
\newcommand{\gevcc}{\ensuremath{{\mathrm{\,Ge\kern -0.1em V\!/}c^2}}\xspace}
\newcommand{\gevgevcccc}{\ensuremath{{\mathrm{\,Ge\kern -0.1em V^2\!/}c^4}}\xspace}
\newcommand{\mevcc}{\ensuremath{{\mathrm{\,Me\kern -0.1em V\!/}c^2}}\xspace}

\usepackage{cite} 
\usepackage{mciteplus}

\begin{document}
\maketitle

\begin{abstract}
	We reanalyze the $\Lambda_b\to p$ transition form factors in the perturbative QCD (PQCD) approach by including higher-twist light-cone distribution amplitudes (LCDAs) of a $\Lambda_b$ baryon and a proton. The previous PQCD evaluation performed decades ago with only the leading-twist $\Lambda_b$ baryon and proton LCDAs gave the form factors, which are two orders of magnitude smaller than indicated by experimental data. We find that the twist-4 $\Lambda_b$ baryon LCDAs and the twist-4 and -5 proton LCDAs contribute dominantly, and the enhanced form factors become consistent with those from lattice QCD and other nonperturbative methods. The estimated branching ratios of the semileptonic decays $\Lambda_b\to p\ell\bar{\nu}_\ell$ and the hadronic decay $\Lambda_b\to p\pi$ are also close to the data. It implies that the $b$ quark mass is not really heavy enough, and higher-power contributions play a crucial role, similar to the observation made in analyses of $B$ meson transition form factors. With the formalism established in this work, we are ready to study various exclusive heavy baryon decays systematically in the PQCD approach.

\end{abstract}
\newpage 
\section{Introduction}

A lot of progresses have been made on probing exclusive $b$-baryon decays with large amount of data collected by LHCb in recent years. $CP$ violation (CPV)
has been established in the $K$, $B$ and $D$ meson systems, but not yet in baryon systems. Therefore, exploring baryon CPV is one of the most important missions in both experimental and theoretical flavor physics. An evidence of CPV has been attained at the confidence level of $3\sigma$ in the $\Lb\to p \pip\pim\pim$ decay~\cite{LHCb:2016yco}. Though CPV is not observed in other modes~\cite{LHCb:2018fly,LHCb:2018fpt,LHCb:2019jyj}, the experimental precision has reached the percent level, such as $A_{CP}(\Lb\to p\Km)=(-2.0\pm1.3\pm1.9)\%$ and $A_{CP}(\Lb\to p\pim)=(-3.5\pm1.7\pm2.0)\%$~\cite{LHCb:2018fly}. The above progresses motivate theoretical investigations on baryon CPV to a similar precision. A QCD-inspired formalism is definitely required for predicting CPV in heavy baryon decays, to which relative strong phases among various amplitudes are the key ingredient. Nevertheless, such a well-developed QCD method has not been available currently.

Several potential frameworks have been proposed for studies of hadronic heavy hadron decays, which include the effective theories such as the heavy quark effective theory (HQET) ~\cite{Mannel:1990vg,Hussain:1992rb,Isgur:1990pm,Georgi:1990cx} and the soft-collinear effective theory (SCET)~\cite{Bauer:2000yr,Bauer:2001yt}, the factorization approaches such as the QCD factorization (QCDF) based on the collinear factorization~\cite{Beneke:1999br,Beneke:2000ry,Beneke:2001ev,Beneke:2003zv} and the perturbative QCD approach (PQCD) based on the $k_T$ factorization~\cite{Keum:2000wi,Lu:2000em,Keum:2000ph}, and phenomenological methods such as topological diagrammatic approaches~\cite{Chau:1987tk,Cheng:2010ry,Li:2012cfa}, final-state interactions \cite{Cheng:2004ru,Yu:2017zst,Han:2021azw} and flavor symmetry analyses \cite{Savage:1989ub,He:2018php,Wang:2020gmn}. The above formalisms have been applied to heavy meson decays extensively, but applications to heavy baryon decays, especially to hadronic decays, are still limited. The generalized factorization assumption has been employed to estimate branching ratios of numerous $b$-baryon decays~\cite{Hsiao:2014mua,Hsiao:2017tif,Geng:2021nkl}. As to QCD-inspired methods, the QCDF approach was applied to $\Lambda_b$ baryon decays under the diquark approximation \cite{Zhu:2016bra}, and the $\Lambda_b\to pK$ and $p\pi$ branching ratios were calculated in the PQCD approach~\cite{Lu:2009cm}, but with the results being several times smaller than experimental data.  

It has been known from global fits to $B$ meson decay data~\cite{Cheng:2014rfa,Qin:2021tve} that  nonfactorizable contributions are crucial for color-suppressed tree-dominated modes, and that the $W$-exchange and penguin-annihilation amplitudes generate large strong phases. The above contributions cannot be computed unambiguously in the other frameworks, such as the factorization assumption and the QCDF approach, but can be in the PQCD approach~\cite{Keum:2000wi,Lu:2000em,Keum:2000ph}. For comparisons of these theoretical methods and their phenomenological impacts, refer to~\cite{Li:2003yj,Li:2007bpa}. It is the reason why the CPV in, for instance, the $B^0\to K^+\pi^-$ and $B^0\to \pi^+\pi^-$ modes~\cite{Belle:2004mad,BaBar:2004gyj,Belle:2004nch}, has been predicted successfully in~\cite{Keum:2000wi,Lu:2000em,Keum:2000ph}. In fact, the PQCD approach has demonstrated a unique power for predicting CPV in two-body hadronic $B$ meson decays. $b$-baryon decays involve more $W$-exchange and penguin-annihilation diagrams~\cite{Han:2021azw,Leibovich:2003tw}, whose PQCD evaluation is feasible in principle. It is our motivation to examine the applicability of the PQCD formalism to exclusive heavy baryon decays in this work.


Baryonic transitions were firstly investigated in the PQCD approach in Ref.~\cite{Li:1992ce}, where the proton Dirac form factors at large momentum transfer were derived. This framework was then extended to studies of heavy-to-light baryonic transition form factors, which are essential inputs to exclusive processes like the semileptonic decays $\Lambda_b \to p\ell\bar{\nu}$~\cite{Shih:1998pb} and $\Lambda_b \to \Lambda_c \ell \bar{\nu}$ at large recoil~\cite{Shih:1999yh,Guo:2005qa}, the radiative decay $\Lambda_b \to \Lambda \gamma$~\cite{He:2006ud}, and the two-body hadronic decays $\Lambda_b \to \Lambda J/\Psi$~\cite{Chou:2001bn}, $\Lambda_b\to p\pi,\ pK$~\cite{Lu:2009cm} and $\Lambda_b\to \Lambda_c\pi,\Lambda_c K$~\cite{Zhang:2022iun}. Only the leading-twist light-cone distribution amplitudes (LCDAs) were considered in the factorization formulas for all the above $\Lambda_b$ baryon decays. It was noticed that the factorizable contributions to two-body hadronic decays are unreasonably smaller than the nonfactorizable ones, and the predicted $\Lambda_b\to p K^-$ branching ratio is several times lower than the measured value~\cite{Lu:2009cm}. In another word, the $\Lambda_b\to p$ transition form factors are down by about two orders of magnitude compared to those from nonperturbative methods in the literature: the PQCD approach gave the $\Lambda_b\to p$ form factor at the maximal recoil $f_1=(2.2^{+0.8}_{-0.5})\times10^{-3}$ in~\cite{Lu:2009cm} and $2.3\times10^{-3}$ in~\cite{Shih:1998pb}, while lattice QCD yielded $f_1=0.22\pm0.08$ \cite{Detmold:2015aaa}. To verify the applicability of PQCD, we should first resolve the difficulty appearing in the $\Lambda_b\to p$ transition form factors. 

The contribution to a heavy-to-light mesonic transition amplitude is divided into two pieces at leading power in the QCDF approach, the nonfactorziable soft form factor and the factorizable hard spectator contribution. The latter is calculated in a perturbation theory, and the former can only be handled in nonperturbative methods, such as lattice QCD~\cite{Detmold:2015aaa,Zhang:2021oja}, QCD sum rules (QSR)~\cite{Huang:1998rq}, light-cone sum rules (LCSR) \cite{Wang:2009hra,Wang:2015ndk}, light-front quark model~\cite{Jaus:1999zv,Wang:2017mqp,Li:2021qod}, etc.. In the baryonic case it has been proved in the SCET~\cite{Wang:2011uv} that the leading-power contribution is completely factorizable due to the absence of an endpoint singularity in the collinear factorization. The corresponding diagrams contain at least two hard-collinear gluon exchanges, so that the leading-power contribution is suppressed by ${\cal O}(\alpha_s^2)$ in the strong coupling. On the contrary, a soft contribution to the baryonic form factor is power-suppressed in the SCET~\cite{Wang:2011uv}, but not down by $\alpha_s$, which turns out to be numerically important. Taking the $\Lambda_b\to\Lambda$ transition form factor $\xi_\Lambda$ at maximal recoil as an example, one got $\xi_\Lambda=-0.012^{+0.009}_{-0.023}$ from the leading-power contribution~\cite{Wang:2011uv}, and $\xi_\Lambda=0.38$ from the SCET sum rules~\cite{Feldmann:2011xf}. It implies that QCD dynamics is quite different between the mesonic and baryonic decays, and that the power suppression from the $b$ quark mass may not be effective.  

The soft form factor for a heavy-to-light baryonic transition is nonfactorizable in the collinear factorization, because an endpoint singularity from small parton momentum fractions will be developed, if the soft form factor is expressed as a convolution of a hard kernel and the baryon LCDAs beyond the leading twist. The PQCD approach based on the $k_T$ factorization, in which parton transverse momenta are kept to avoid the endpoint singularity~\cite{Li:2014xda,JHEP02-008}, provides a new set of power counting rules~\cite{Li:2003yj,Liu:2020upy}. The $k_T$ resummation is demanded to organize the large logarithms owing to the introduction of the additional scales $k_T$. The resultant Sudakov factors suppress the long-distance contributions now characterized by large $b$ with $b$ being the variables conjugate to $k_T$. Once the endpoint singularity is smeared, the higher-twist contribution to the heavy-to-light form factor is regarded as being factorizable and calculable in the PQCD formalism. The factorizable contribution then picks up the additional piece from the higher-twist LCDAs under the Sudakov suppression, such that the previous leading-twist PQCD results can be significantly enhanced.


The above discussion suggests that the contributions from higher-twist LCDAs are factorizable in the $k_T$ factorization, and their inclusion may increase the much smaller $\Lambda_b\to p$ form factors at leading power~\cite{Lu:2009cm,Shih:1998pb}. Here we will analyze these form factors in the fast recoil region by including the $\Lambda_b$ baryon LCDAs up to twist 4 and the proton LCDAs up to twist 6 in the PQCD approach. It will be shown that the $\Lambda_b\to p$ form factors become comparable to those derived from other nonperturbative methods and indicated by experimental data, when the above higher-twist LCDAs are taken into account. In particular, the convolution with the twist-4 $\Lambda_b$ baryon LCDAs and the twist-4 and -5 proton LCDAs give the dominant contributions. We then estimate the branching ratios of the semileptonic decays $\Lambda_b\to p\ell\bar{\nu}_\ell$ with the leptons $\ell=e,\mu,\tau$ by extrapolating the form factors at large recoil to the whole kinematic range. The agreement of our results with data encourages the generalization of the established PQCD formalism to more complicated two-body hadronic heavy baryon decays.



The remainder of this paper is organized as follows. The PQCD framework for computing the $\Lambda_b\to p$ transition form factors is explained in Sec.~\ref{sec:framework}, where the $\Lambda_b$ baryon and proton LCDAs of various twists are defined. The numerical outcomes for the $\Lambda_b\to p$ form factors and for the differential widths of the semileptonic decays $\Lambda_b\to p\ell\bar{\nu}_\ell$ are presented and discussed in Sec.~\ref{sec:results}. We also make the preliminary prediction for the $\Lambda_b\to p\pi$ branching ratio based on the naive factorization assumption. The last section contains the conclusion. The explicit expressions for the factorization formulas together with the hard scales involved in various diagrams are collected in the Appendix.
\section{Theoretical framework}\label{sec:framework}

\subsection{$k_T$ factorization}
The $\Lambda_b \to p$ transition form factors are defined via the matrix element of the $V-A$ current \cite{Mannel:1990vg},
\begin{align}
\langle P(p^\prime,s^\prime) | \bar{u}\gamma_\mu(1-\gamma_5)b | \Lambda_b(p,s) \rangle =\nonumber& \overline{N}(p^\prime,s^\prime) (f_1 \gamma_\mu - i f_2 \sigma_{\mu\nu} q^\nu + f_3 q_\mu)\Lambda_b(p,s)\nonumber\\
& -\overline{N}(p^\prime,s^\prime) (g_1 \gamma_\mu - i g_2 \sigma_{\mu\nu} q^\nu + g_3 q_\mu)\gamma_5\Lambda_b(p,s),
\label{eq:FF}
\end{align}
where $\sigma_{\mu\nu}=i[\gamma_\mu,\gamma_\nu]/2$, and $\Lambda_b(p,s)$ ($N(p^\prime,s^\prime)$) is the spinor of the $\Lambda_b$ baryon (proton) with the momentum $p$ ($p^\prime$) and the spin $s$ ($s^\prime$). The form factors $f_i$ and $g_i$ depend on the invariant mass squared of the lepton pair $q^2$ with $q_\mu=p_\mu-p^\prime_\mu$. We work in the rest frame of the $\Lambda_b$ baryon, and parameterize the $\Lambda_b$ baryon and proton momenta in the light-cone coordinates as
\begin{eqnarray}
p=\frac{m_{\Lambda_b}}{\sqrt{2}}(1,1,\textbf{0}),\;\;\;\;
p^\prime=\frac{m_{\Lambda_b}}{\sqrt{2}}(\eta_1,\eta_2,\textbf{0}),
\end{eqnarray}
with the large component $\eta_1\sim {\cal O}(1)$ and the small component $\eta_2\sim {\cal O}(m_p^2/m_{\Lambda_b}^2)$, $m_{\Lambda_b}$ ($m_p$) being the $\Lambda_b$ baryon (proton) mass. Namely, the fast recoiled proton has been assumed to move approximately in the plus direction. The invariant mass squared of the lepton pair is then given, in terms of $\eta_1$ and $\eta_2$, by $q^2=m^2_{\Lambda_b}(1-\eta_1)(1-\eta_2)$. 

\begin{figure}[htbp]
	\includegraphics[scale=0.35]{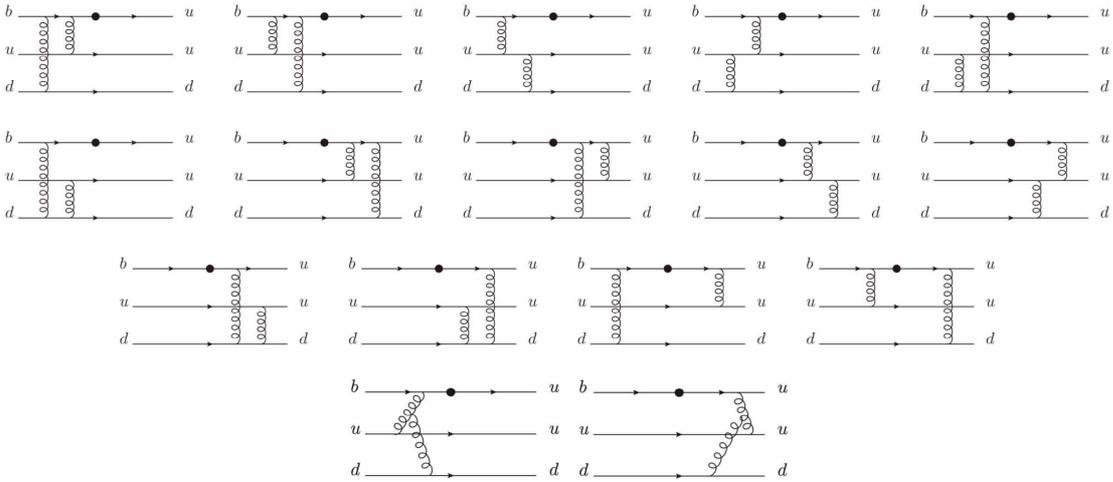}
	\caption{Feynman diagrams for the $\Lambda_b\to p$ transition form factors, where the black dots denote the weak interaction vertices. These diagrams are labelled by $D_1$, $D_2$,... and $D_{16}$ in sequence in the text.}
	\label{fig:feynman}
\end{figure}

As stated before, the transverse momenta of the valence quarks are retained in the PQCD approach based on the $k_T$ factorization. We thus choose the partonic momenta as
\begin{align}
k_1=&({m_{\Lambda_b}\over\sqrt{2}},{x_1m_{\Lambda_b}\over\sqrt{2}},\textbf{k}_{1T}),\ \ \ \ k_1^\prime=(x_1^\prime{\eta_1m_{\Lambda_b}\over\sqrt{2}},0,\textbf{k}_{1T}^\prime),\nonumber\\
k_2=&(\,\,\,0,\,\,\,\,\,\,{x_2m_{\Lambda_b}\over\sqrt{2}},\textbf{k}_{2T}),\ \ \ \ k_2^\prime=(x_2^\prime{\eta_1m_{\Lambda_b}\over\sqrt{2}},0,\textbf{k}_{2T}^\prime),\nonumber\\
k_3=&(\,\,\,0,\,\,\,\,\,\,{x_3m_{\Lambda_b}\over\sqrt{2}},\textbf{k}_{3T}),\ \ \ \ k_3^\prime=(x_3^\prime{\eta_1m_{\Lambda_b}\over\sqrt{2}},0,\textbf{k}_{3T}^\prime),
\end{align}
where $k_1$ is the $b$ quark momentum, $k_2$ ($k_3$) and $k_2^\prime$ ($k_3^\prime$) are the spectator $u$ ($d$) quark momenta in the $\Lambda_b$ baryon and the proton, respectively, and $x_i,k_{iT}$ and $x_i^\prime,k_{iT}^\prime$ denote the corresponding light-cone momentum fractions and the transverse momenta. Note that $x_1$ is of ${\cal O}(m_b^2/m_{\Lambda_b}^2)$, $m_b$ being $b$ quark mass, in order for the $b$ quark to be off-shell by $k_1^2\approx -k_{1T}^2$, as required by the $k_T$ factorization. The dominant plus components of $k_i^\prime$ are kept, and the minus components of $k_i$ for the soft light quarks are selected by their inner products with $k_i^\prime$, which appear in the hard kernels for the $\Lambda_b\to p$ form factors. Since the two soft quarks in the $\Lambda_b$ baryon need to turn into the energetic quarks in the proton, at least two hard gluons are exchanged as shown in the leading-order Feynman diagrams in Fig.~\ref{fig:feynman}. That is, the $\Lambda_b \to p$ decay amplitudes start at  ${\cal O}(\alpha_s^2)$ in the PQCD approach. 

The $\Lambda_b\to p$ transition amplitude is formulated in the PQCD approach as~\cite{Lu:2009cm}
\begin{eqnarray}
\mathcal{A}=\Psi_{\Lambda_b}(x_i,b_i,\mu)\otimes H(x_i,b_i,x_i^{\prime},b_i^{\prime},\mu)\otimes \Psi_{P}(x_i^{\prime},b_i^{\prime},\mu),
\end{eqnarray}
where the hard kernel $H$ is derived from the diagrams in Fig.~\ref{fig:feynman}, $\Psi_{\Lambda_b}$ and $\Psi_{P}$ stand for the  $\Lambda_b$ baryon and proton wave functions, respectively, and $\mu$ is the factorization scale. The symbol $\otimes$ represents the convolution in the momentum fractions and in the impact parameters $b_i$ and $b_i^{\prime}$, which are conjugated to the corresponding transverse momenta. With the parton transverse degrees of freedom being included, the above $k_T$ factorization formula holds for higher-twist contributions due to the absence of endpoint singularities. At the same time, the large double logarithms $\alpha_s\ln^2 (m_{\Lambda_b}b)$ are produced from radiative corrections to the baryon wave functions, which must be summed to all orders in $\alpha_s$ to improve the convergence of perturbation expansion. The $k_T$ resummation is thus applied to the baryon wave functions to extract the Sudakov factors~\cite{Lu:2009cm,Bolz:1994hb,Kundu:1998gv} as in its application to meson wave functions~\cite{Li:1994cka,Li:1994iu}. The Sudakov factor, which decreases fast with $b$ and vanishes at $b=1/\Lambda_{\rm QCD}$, $\Lambda_{\rm QCD}$ being the QCD scale, is expected to effectively suppress the long-distance contributions from the large $b$ regions. We also implement the renormalization-group evolution in $\mu$ up to a hard scale $t$, which is set to the maximum of all scales involved in the factorization formula~\cite{Lu:2009cm}. The necessary constraint on the hard scale, $t\ge 1$ GeV, is imposed, because it should not go below the initial scale, at which models for the baryon wave functions are defined.

We then arrive at an factorization formula improved by the resummation and renormalization-group evolution, 
\begin{eqnarray}\label{eq:PQCD}
\mathcal{A}=\psi_{\Lambda_b}(x_i,b_i)\otimes H(x_i,b_i,x_i^{\prime},b_i^{\prime},t)\otimes \psi_{P}(x_i^{\prime},b_i^{\prime})\exp\left[-S_{\Lambda_b}(t)-S_p(t)\right],
\end{eqnarray}
where the $\Lambda_b$ baryon (proton) wave function $\psi_{\Lambda_b}$ ($\psi_{P}$) is obtained by factorizing the extrinsic impact-parameter dependence of the wave function $\Psi_{\Lambda_b}$ ($\Psi_{P}$) into the total exponential factor $\exp[-S_{\Lambda_b}(t)]$ ($\exp[-S_p(t)]$)~\cite{Lu:2009cm}. The remaining impact-parameter dependencies of $\psi_{\Lambda_b}$ and $\psi_{P}$ are intrinsic. We point out that another type of double logarithms $\alpha_s\ln^2x_i$, appearing in the hard kernel, will also become crucial as the endpoint regions dominate. The threshold resummation that sums this type of double logarithms to all orders leads to the jet function (or the threshold Sudakov factor) $S_t(x_i)$~\cite{Li:2001ay,Zhang:2020qaz}, which can further improve the perturbative expansion. The jet function  $S_t(x_i)$, extracted from the hard kernel, is process-dependent. Because a systematic derivation of $S_t(x_i)$ for baryonic decays goes beyond the scope of this paper, we will naively set $S_t(x_i)=1$ below, and investigate its effect in the future. The higher-twist $\Lambda_b$ baryon and proton wave functions and the corresponding power-suppressed contributions will be analyzed quantitatively based on Eq.~(\ref{eq:PQCD}). The factorization formulas for the $\Lambda_b \to p$ form factors from the diagrams in Fig.~\ref{fig:feynman} and the involved hard scales are provided in Appendix~\ref{app:amplitude}. It will be seen that the predictions for these form factors as well as the relevant observables in the PQCD formalism are substantially enhanced.

\subsection{Light-cone distribution amplitudes of baryons}\label{sec:LCDAs}
Hadronic wave functions are universal nonperturbative inputs to a factorization theorem, which need to be specified before predictions for an exclusive QCD process are made. The explicit form of a wave function, being a three-dimension object, may be very complicated. As a usual practice, one assumes that a wave function is factorized into a product of a momentum-fraction-dependent part and a transverse-momentum-dependent part. The former corresponds to a LCDA for the collinear factorization, and the latter has been parameterized as a simple Gaussian type function. This assumption has been widely adopted in applications of the $k_T$ factorization to exclusive processes. Since the transverse-momentum, i.e., impact-parameter dependencies of the baryon wave functions $\psi_{\Lambda_b}$ and $\psi_{P}$ are still not well constrained, we will neglect them in our numerical studies.

\subsubsection{$\Lambda_b$ baryon light-cone distribution amplitudes}\label{sec:LbDAs}
The $\Lambda_b$ baryon LCDAs are defined by the matrix elements of nonlocal operators sandwiched between the vacuum and the $\Lambda_b$ baryon state, whose general Lorentz structures can be found in Refs.~\cite{Ball:2008fw,Bell:2013tfa,Wang:2015ndk,Ali:2012zza}. We start with the momentum-space projector
\begin{equation}
(Y_{\Lambda_b})_{\alpha\beta\gamma}(x_i,\mu)=\frac{1}{8N_c}\Big\{f_{\Lambda_b}^{(1)}(\mu)[M_1(x_2,x_3)\gamma_5C^T]_{\gamma\beta}+f_{\Lambda_b}^{(2)}(\mu)[M_2(x_2,x_3)\gamma_5C^T]_{\gamma\beta}\Big\}[\Lambda_b(p)]_\alpha,\label{pro}
\end{equation}
where $N_c$ is the number of colors, the normalization constants $f_{\Lambda_b}^{(1)} \approx f_{\Lambda_b}^{(2)}\equiv f_{\Lambda_b}= 0.021\pm 0.004$ GeV$^3$, which are consistent with $f_{\Lambda_b}= 0.022\pm 0.001$ GeV$^3$ quoted from the leading-order sum rule calculation~\cite{Groote:1997yr}, $C$ represents the charge conjugation matrix, and $\Lambda_b(p)$ is the $\Lambda_b$ baryon spinor. Note that the normalization constants have been derived in diagonal, non-diagonal and mixed sum rules at the leading-order and next-to-leading-order levels in~\cite{Groote:1997yr}. Because the equality $f_{\Lambda_b}^{(1)}\approx f_{\Lambda_b}^{(2)}$ assumed above is not guaranteed in non-diagonal and mixed sum rules~\cite{Groote:1997yr}, and invalidated under next-to-leading-order corrections~\cite{Groote:1996em}, it is more consistent to adopt the result for $f_{\Lambda_b}^{(1,2)}$ from the leading-order diagonal sum rules here. The terms containing the derivatives with respect to the transverse momenta of the soft light quarks have been ignored in Eq.~(\ref{pro}). Their contributions are expected to be tiny, similar to what was observed in the PQCD analysis of $B$ meson transition form factors~\cite{Li:2012nk}. 

The remaining parts of the projector in Eq.~(\ref{pro}) are expressed as 
\begin{align}
M_1(x_2,x_3)=&\frac{\slashed{\bar{n}}\slashed{n}}{4}\psi_3^{+-}(x_2,x_3)+\frac{\slashed{n}\slashed{\bar{n}}}{4}\psi_3^{-+}(x_2,x_3),\\
M_2(x_2,x_3)=&\frac{\slashed{n}}{\sqrt{2}}\psi_2(x_2,x_3)+\frac{\slashed{\bar{n}}}{\sqrt{2}}\psi_4(x_2,x_3),
\end{align}
where the two light-cone vectors $n=(1,0,\textbf{0})$ and $\bar{n}=(0,1,\textbf{0})$ satisfy $n\cdot\bar{n}=1$. Various models for the $\Lambda_b$ baryon LCDAs $\psi_2$, $\psi_3^{+-}$, $\psi_3^{-+}$ and $\psi_4$ have been proposed in Refs.~\cite{Ball:2008fw,Bell:2013tfa,Ali:2012zza}. Viewing the obvious difference among these models, we will investigate the contributions to the $\Lambda_b\to p$ form factors from all of them for completeness:
\begin{itemize}
	\item Gegenbauer-1~\cite{Ball:2008fw}, which was obtained by taking into account only the leading-order perturbative contribution to the associated QCD sum rules,
	\begin{align}
	\psi_2(x_2,x_3)=&m_{\Lambda_b}^4x_2x_3\left[\frac{1}{\epsilon_0^4}e^{-m_{\Lambda_b}(x_2+x_3)/\epsilon_0}+a_2C_2^{3/2}(\frac{x_2-x_3}{x_2+x_3})\frac{1}{\epsilon_1^4}e^{-m_{\Lambda_b}(x_2+x_3)/\epsilon_1}\right],\nonumber\\
	\psi_3^{+-}(x_2,x_3)=&\frac{2m_{\Lambda_b}^3x_2}{\epsilon_3^3}e^{-m_{\Lambda_b}(x_2+x_3)/\epsilon_3},\nonumber\\
	\psi_3^{-+}(x_2,x_3)=&\frac{2m_{\Lambda_b}^3x_3}{\epsilon_3^3}e^{-m_{\Lambda_b}(x_2+x_3)/\epsilon_3},\nonumber\\
	\psi_4(x_2,x_3)=&\frac{5}{\mathcal{N}}m_{\Lambda_b}^2\int_{m_{\Lambda_b}(x_2+x_3)/2}^{s_0}ds e^{-s/\tau} (s-m_{\Lambda_b}(x_2+x_3)/2)^3,
	\end{align}
	with the Gegenbauer moment $a_2=0.333_{-0.333}^{+0.250}$, the Gegenbauer polynomial $C_2^{3/2}(x)=3(5x^2-1)/2$, the parameters $\epsilon_0=200_{-60}^{+130}$ MeV, $\epsilon_1=650_{-300}^{+650}$ MeV and $\epsilon_3=230\pm 60$ MeV, the Borel mass 0.4 GeV $<\tau<$ 0.8 GeV, the continuum threshold $s_0=1.2$ GeV and the constant $\mathcal{N}=\int_{0}^{s_0}dss^5e^{-s/\tau}$.
	
	\item Gegenbauer-2~\cite{Ali:2012zza}, which was formulated in the heavy quark limit with the moments being derived in QCD sum rules,
	{\footnotesize
		\begin{align}
		\psi_2(x_2,x_3)=&m_{\Lambda_b}^4x_2x_3\left(\frac{a_0^{(2)}}{{\epsilon_0^{(2)}}^4}C_0^{3/2}(\frac{x_2-x_3}{x_2+x_3})e^{-m_{\Lambda_b}(x_2+x_3)/\epsilon_0^{(2)}} + \frac{a_2^{(2)}}{{\epsilon_2^{(2)}}^4}C_2^{3/2}(\frac{x_2-x_3}{x_2+x_3})e^{-m_{\Lambda_b}(x_2+x_3)/\epsilon_2^{(2)}}\right),\nonumber\\
		\psi_3^{+-}(x_2,x_3)=&m_{\Lambda_b}^3(x_2+x_3)\left[\frac{a_0^{(3)}}{{\epsilon_0^{(3)}}^3}C_0^{1/2}(\frac{x_2-x_3}{x_2+x_3})e^{-m_{\Lambda_b}(x_2+x_3)/\epsilon_0^{(3)}} + \frac{a_2^{(3)}}{{\epsilon_2^{(3)}}^3}C_2^{1/2}(\frac{x_2-x_3}{x_2+x_3})e^{-m_{\Lambda_b}(x_2+x_3)/\epsilon_2^{(3)}} \right],\nonumber\\
		&+m_{\Lambda_b}^3(x_2+x_3)\left[\frac{b_1^{(3)}}{{\eta_1^{(3)}}^3}C_1^{1/2}(\frac{x_2-x_3}{x_2+x_3})e^{-m_{\Lambda_b}(x_2+x_3)/\eta_1^{(3)}} + \frac{b_3^{(3)}}{{\eta_3^{(3)}}^3}C_2^{1/2}(\frac{x_2-x_3}{x_2+x_3})e^{-m_{\Lambda_b}(x_2+x_3)/\eta_3^{(3)}}\right],\nonumber\\
		\psi_3^{-+}(x_2,x_3)=&m_{\Lambda_b}^3(x_2+x_3)\left[\frac{a_0^{(3)}}{{\epsilon_0^{(3)}}^3}C_0^{1/2}(\frac{x_2-x_3}{x_2+x_3})e^{-m_{\Lambda_b}(x_2+x_3)/\epsilon_0^{(3)}} + \frac{a_2^{(3)}}{{\epsilon_2^{(3)}}^3}C_2^{1/2}(\frac{x_2-x_3}{x_2+x_3})e^{-m_{\Lambda_b}(x_2+x_3)/\epsilon_2^{(3)}} \right],\nonumber\\
		&-m_{\Lambda_b}^3(x_2+x_3)\left[\frac{b_1^{(3)}}{{\eta_1^{(3)}}^3}C_1^{1/2}(\frac{x_2-x_3}{x_2+x_3})e^{-m_{\Lambda_b}(x_2+x_3)/\eta_1^{(3)}} + \frac{b_3^{(3)}}{{\eta_3^{(3)}}^3}C_2^{1/2}(\frac{x_2-x_3}{x_2+x_3})e^{-m_{\Lambda_b}(x_2+x_3)/\eta_3^{(3)}}\right],\nonumber\\
		\psi_4(x2,x3)=&m_{\Lambda_b}^2\left(\frac{a_0^{(4)}}{{\epsilon_0^{(4)}}^2}C_0^{1/2}(\frac{x_2-x_3}{x_2+x_3})e^{-m_{\Lambda_b}(x_2+x_3)/\epsilon_0^{(4)}} + \frac{a_2^{(4)}}{{\epsilon_2^{(4)}}^2}C_2^{1/2}(\frac{x_2-x_3}{x_2+x_3})e^{-m_{\Lambda_b}(x_2+x_3)/\epsilon_2^{(4)}}\right),
		\end{align}}
	with the Gegenbauer polynomials $C_0^{3/2}(x)=1$, $C_0^{1/2}(x)=1$, $C_2^{1/2}(x)=(2x^2-1)/2$, and the parameters $a_0^{(2)}=1$, $a_0^{(3)}=1$, $a_0^{(4)}=1$, $a_2^{(2)}=0.391\pm0.279$, $a_2^{(3)}=-0.161_{-0.207}^{+0.108}$, $a_2^{(4)}=-0.541_{-0.09}^{+0.173}$, $b_1^{(3)}=1$, $b_3^{(3)}=-0.24_{-0.147}^{+0.24}$, $\epsilon_0^{(2)}=0.201_{-0.059}^{+0.143}$ GeV, $\epsilon_0^{(3)}=0.232_{-0.056}^{+0.047}$ GeV, $\epsilon_0^{(4)}=0.352_{-0.083}^{+0.067}$ GeV, $\epsilon_2^{(2)}=0.551_{-0.356}^{+\infty}$ GeV, $\epsilon_2^{(3)}=0.055_{-0.02}^{+0.01}$ GeV, $\epsilon_2^{(4)}=0.262_{-0.132}^{+0.116}$ GeV, $\eta_1^{(3)}=0.324_{-0.026}^{+0.054}$ GeV and $\eta_3^{(3)}=0.633\pm0.099$ GeV.
	
	\item Exponential model~\cite{Bell:2013tfa},\\
	\begin{align}
	\psi_2(x_2,x_3)=& \frac{x_2x_3}{\omega_0^4}m^4_{\Lambda_b}e^{-(x_2+x_3)m_{\Lambda_b}/\omega_0},\nonumber\\
	\psi_3^{+-}(x_2,x_3)=& \frac{2x_2}{\omega_0^3}m^3_{\Lambda_b}e^{-(x_2+x_3)m_{\Lambda_b}/\omega_0},\nonumber\\
	\psi_3^{-+}(x_2,x_3)=& \frac{2x_3}{\omega_0^3}m^3_{\Lambda_b}e^{-(x_2+x_3)m_{\Lambda_b}/\omega_0},\nonumber\\
	\psi_4(x_2,x_3)=& \frac{1}{\omega_0^2}m_{\Lambda_b}^2e^{-(x_2+x_3)m_{\Lambda_b}/\omega_0},
	\label{eq:EXP}
	\end{align}
	where $\omega_0=0.4$ GeV 
	measures the average energy of the two light quarks.
	\item Free-parton approximation~\cite{Bell:2013tfa},
	\begin{align}
	\psi_2(x_2,x_3)=& \frac{15x_2x_3m_{\Lambda_b}^4(2\bar{\Lambda}-x_2m_{\Lambda_b}-x_3m_{\Lambda_b})}{4\bar{\Lambda}^5}\Theta(2\bar{\Lambda}-x_2m_{\Lambda_b}-x_3m_{\Lambda_b}),\nonumber\\
	\psi_3^{+-}(x_2,x_3)=&\frac{15x_2m_{\Lambda_b}^3(2\bar{\Lambda}-x_2m_{\Lambda_b}-x_3m_{\Lambda_b})^2}{4\bar{\Lambda}^5}\Theta(2\bar{\Lambda}-x_2m_{\Lambda_b}-x_3m_{\Lambda_b}),\nonumber\\
	\psi_3^{-+}(x_2,x_3)=&\frac{15x_3m_{\Lambda_b}^3(2\bar{\Lambda}-x_2m_{\Lambda_b}-x_3m_{\Lambda_b})^2}{4\bar{\Lambda}^5}\Theta(2\bar{\Lambda}-x_2m_{\Lambda_b}-x_3m_{\Lambda_b}),\nonumber\\
	\psi_4(x_2,x_3)=& \frac{5m_{\Lambda_b}^2(2\bar{\Lambda}-x_2m_{\Lambda_b}-x_3m_{\Lambda_b})^3}{8\bar{\Lambda}^5}\Theta(2\bar{\Lambda}-x_2m_{\Lambda_b}-x_3m_{\Lambda_b}),
	\label{eq:PAR}
	\end{align}
	with the theta function $\Theta$ and the scale $\bar{\Lambda}\equiv (m_{\Lambda_b}-m_b)/2\approx 0.8$ GeV.
\end{itemize}
All the above models for the $\Lambda_b$ baryon  LCDAs obey the normalizations
\begin{align}
&\int_{0}^{1}dx_1dx_2dx_3 \delta(1-x_1-x_2-x_3)\psi_2(x_2,x_3)=1,\nonumber\\
&\int_{0}^{1}dx_1dx_2dx_3 \delta(1-x_1-x_2-x_3)(\psi_3^{+-}(x_2,x_3)+\psi_3^{-+}(x_2,x_3))/4=1,\nonumber\\
&\int_{0}^{1}dx_1dx_2dx_3 \delta(1-x_1-x_2-x_3)\psi_4(x_2,x_3)=1.
\end{align}

In order to compare our results with the previous ones, we quote the simple model for the leading-twist $\Lambda_b$ baryon LCDA proposed in Ref.~\cite{Schlumpf:1992ce},
\begin{equation}
(Y_{\Lambda_b})_{\alpha\beta\gamma}(x_i,\mu)=\frac{f_{\Lambda_b}^\prime}{8\sqrt{2}N_c}[(\slashed{p}+m_{\Lambda_b})\gamma_5C]_{\beta\gamma}[\Lambda_b(p)]_\alpha\psi(x_i,\mu),
\label{eq:simplified}
\end{equation}
\begin{equation}
\psi(x_i)=Nx_1x_2x_3 \exp\left(-\frac{m_{\Lambda_b}^2}{2\beta^2x_1}-\frac{m_l^2}{2\beta^2x_2}-\frac{m_l^2}{2\beta^2x_3}\right),
\label{simmod}\end{equation}
whose Lorentz structure has been simplified under the Bargmann-Wigner equation~\cite{Hussain:1990uu} in the heavy quark limit, such that the spin and orbital degrees of freedom of the light quark system are decoupled. The $\mu$ dependence can be organized into the total exponential factor as stated before, $f_{\Lambda_b}^\prime$ is set to the value  $4.28^{+0.75}_{-0.64}\times 10^{-3}$ $\text{GeV}^2$ the same as in the previous analysis \cite{Lu:2009cm}, $N=6.67\times10^{12}$ is determined by the normalization condition, the shape parameter takes the value $\beta=1.0\pm0.2$ GeV, and $m_l=0.3$ GeV represents the mass of the light degrees of freedom in the $\Lambda_b$ baryon. 

\subsubsection{Proton light-cone distribution amplitudes}
The proton LCDAs with definite twists have been defined in Ref.~\cite{Braun:2000kw}, and the corresponding momentum-space projector is written as
{\small
	\begin{align}\label{eq:proton DAs}
	(\overline{Y}_P)_{\alpha\beta\gamma}&(x_i^\prime,\mu)=\frac{1}{8\sqrt{2}N_c}\Big\{
	S_1 m_p C_{\beta\alpha} (\bar{N}^+ \gamma_5)_\gamma + S_2 m_p C_{\beta\alpha} (\bar{N}^- \gamma_5)_\gamma + P_1 m_p (C\gamma_5)_{\beta\alpha} \bar{N}^+_\gamma\nonumber\\
	&+ P_2 m_p (C\gamma_5)_{\beta\alpha} \bar{N}^-_\gamma+ V_1 (C\slashed{P})_{\beta\alpha} (\bar{N}^+\gamma_5)_\gamma + V_2 (C\slashed{P})_{\beta\alpha} (\bar{N}^-\gamma_5)_\gamma\nonumber\\
	&+ V_3 \frac{m_p}{2} (C\gamma_\perp)_{\beta\alpha}(\bar{N}^+\gamma_5\gamma^\perp)_\gamma+ V_4 \frac{m_p}{2} (C\gamma_\perp)_{\beta\alpha}(\bar{N}^-\gamma_5\gamma^\perp)_\gamma + V_5\frac{m_p^2}{2Pz} (C\slashed{z})_{\beta\alpha}(\bar{N}^+\gamma_5)_\gamma\nonumber\\
	&+ V_6\frac{m_p^2}{2Pz} (C\slashed{z})_{\beta\alpha}(\bar{N}^-\gamma_5)_\gamma+ A_1 (C\gamma_5\slashed{P})_{\beta\alpha} (\bar{N}^+)_\gamma+ A_2 (C\gamma_5\slashed{P})_{\beta\alpha} (\bar{N}^-)_\gamma\nonumber\\
	&+ A_3 \frac{m_p}{2} (C\gamma_5\gamma_\perp)_{\beta\alpha}(\bar{N}^+\gamma^\perp)_\gamma+ A_4 \frac{m_p}{2} (C\gamma_5\gamma_\perp)_{\beta\alpha}(\bar{N}^-\gamma^\perp)_\gamma+ A_5\frac{m_p^2}{2Pz} (C\gamma_5\slashed{z})_{\beta\alpha}(\bar{N}^+)_\gamma \nonumber\\
	&+ A_6\frac{m_p^2}{2Pz} (C\gamma_5\slashed{z})_{\beta\alpha}(\bar{N}^-)_\gamma- T_1 (iC\sigma_{\perp P})_{\beta\alpha}(\bar{N}^+\gamma_5\gamma^\perp)_\gamma- T_2 (iC\sigma_{\perp P})_{\beta\alpha}(\bar{N}^-\gamma_5\gamma^\perp)_\gamma \nonumber\\
	&- T_3 \frac{m_p}{Pz}(iC\sigma_{Pz})_{\beta\alpha}(\bar{N}^+\gamma_5)_\gamma- T_4 \frac{m_p}{Pz}(iC\sigma_{zP})_{\beta\alpha}(\bar{N}^-\gamma_5)_\gamma - T_5\frac{m_p^2}{2Pz}(iC\sigma_{\perp z})_{\beta\alpha}(\bar{N}^+\gamma_5\gamma^\perp)_\gamma \nonumber\\
	&- T_6\frac{m_p^2}{2Pz}(iC\sigma_{\perp z})_{\beta\alpha}(\bar{N}^-\gamma_5\gamma^\perp)_\gamma+ T_7\frac{m_p}{2}(C\sigma_{\perp\perp^\prime})_{\beta\alpha}(\bar{N}^+\gamma_5\sigma^{\perp\perp^\prime})_\gamma \nonumber\\
	&+ T_8\frac{m_p}{2}(C\sigma_{\perp\perp^\prime})_{\beta\alpha}(\bar{N}^-\gamma_5\sigma^{\perp\perp^\prime})_\gamma
	\Big\},
	\end{align}}
with the proton mass $m_p=0.938$ GeV. The light-like vector $P$ can be decomposed into
\begin{equation}
P_\mu=p_\mu^\prime-\frac{1}{2}z_\mu\frac{m_p^2}{Pz},
\end{equation}
where $p^\prime$ is the proton momentum, and $z$ is a light-like vector with $z^2=0$. We have adopted the shorthand notations $\sigma_{Pz}=\sigma^{\nu\mu}P_\nu z_\mu$, and $\bar{N}^+=\bar{N}\slashed{z}\slashed{P}/(2Pz)$ and $\bar{N}^-=\bar{N}\slashed{P}\slashed{z}/(2Pz)$ for the ``large" and ``small" components of the proton spinor $\overline{N}$, respectively. The symbol $\perp$ denotes the projection perpendicular to $z$ or $P$, and the contraction $\gamma_\perp\gamma^\perp$ means $\gamma_\perp\gamma^\perp=\gamma^\mu g_{\mu\nu}^\perp\gamma^\nu$ with $g_{\mu\nu}^\perp=g_{\mu\nu}-(P_\mu z_\nu+z_\mu P_\nu)/Pz$. The twist classification of the LCDAs $V_i$, $A_i$, $T_i$, $S_i$ and $P_i$ is specified in Table~\ref{table: proton twist classification}, and their explicit expressions are listed below:
\begin{itemize}
	\item Twist-3 LCDAs\\
	{\footnotesize
		\begin{align}
		V_1(x_i)=&120x_1x_2x_3[\phi_3^0+\phi_3^+(1-3x_3)],\\
		A_1(x_i)=&120x_1x_2x_3(x_2-x_1)\phi_3^-,\\
		T_1(x_i)=&120x_1x_2x_3[\phi_3^0+\frac{1}{2}(\phi_3^--\phi_3^+)(1-3x_3)].
		\end{align}}
	\item Twist-4 LCDAs\\
	{\footnotesize
		\begin{align}
		V_2(x_i)=&24x_1x_2[\phi_4^0+\phi_4^+(1-5x_3)],\\
		V_3(x_i)=&12x_3[\psi_4^0(1-x_3)+\psi_4^-(x_1^2+x_2^2-x_3(1-x_3))+\psi_4^+(1-x_3-10x_1x_2)],\\
		A_2(x_i)=&24x_1x_2(x_2-x_1)\phi_4^-,\\
		A_3(x_i)=&12x_3(x_2-x_1)[(\psi_4^0+\psi_4^+)+\psi_4^-(1-2x_3)],\\
		T_2(x_i)=&24x_1x_2[\xi_4^0+\xi_4^+(1-5x_3)],\\
		T_3(x_i)=&6x_3[(\xi_4^0+\phi_4^0+\psi_4^0)(1-x_3)+(\xi_4^-+\phi_4^--\psi_4^-)(x_1^2+x_2^2-x_3(1-x_3))\nonumber\\
		&+(\xi_4^++\phi_4^++\psi_4^+)(1-x_3-10x_1x_2)],\\
		T_7(x_i)=&6x_3[(-\xi_4^0+\phi_4^0+\psi_4^0)(1-x_3)+(-\xi_4^-+\phi_4^--\psi_4^-)(x_1^2+x_2^2-x_3(1-x_3))\nonumber\\
		&+(-\xi_4^++\phi_4^++\psi_4^+)(1-x_3-10x_1x_2)],\\
		S_1(x_i)=&6x_3(x_2-x_1)[(\xi_4^0+\phi_4^0+\psi_4^0+\xi_4^++\phi_4^++\psi_4^+)+(\xi_4^-+\phi_4^--\psi_4^-)(1-2x_3)],\\
		P_1(x_i)=&6x_3(x_2-x_1)[(\xi_4^0-\phi_4^0-\psi_4^0+\xi_4^+-\phi_4^+-\psi_4^+)+(\xi_4^--\phi_4^-+\psi_4^-)(1-2x_3)].
		\end{align}}
	\item Twist-5 LCDAs\\
	{\footnotesize
		\begin{align}
		V_4(x_i)=&3[\psi_5^0(1-x_3)+\psi_5^-(2x_1x_2-x_3(1-x_3))+\psi_5^+(1-x_3-2(x_1^2+x_2^2))],\\
		V_5(x_i)=&6x_3[\phi_5^0+\phi_5^+(1-2x_3)],\\
		A_4(x_i)=&3(x_2-x_1)[-\psi_5^0+\psi_5^-x_3+\psi_5^+(1-2x_3)],\\
		A_5(x_i)=&6x_3(x_2-x_1)\phi_5^-,\\
		T_4(x_i)=&\frac{3}{2}[(\xi_5^0+\psi_5^0+\phi_5^0)(1-x_3)+(\xi_5^-+\phi_5^--\psi_5^-)(2x_1x_2-x_3(1-x_3))\nonumber\\
		&+(\xi_5^++\phi_5^++\psi_5^+)(1-x_3-2(x_1^2+x_2^2))],\\
		T_5(x_i)=&6x_3[\xi_5^0+\xi_5^+(1-2x_3)],\\
		T_8(x_i)=&\frac{3}{2}[(\psi_5^0+\phi_5^0-\xi_5^0)(1-x_3)+(\phi_5^--\phi_5^--\xi_5^-)(2x_1x_2-x_3(1-x_3))\nonumber\\
		&+(\phi_5^++\phi_5^+-\xi_5^+)(\mu)(1-x_3-2(x_1^2+x_2^2))],\\
		S_2(x_i)=&\frac{3}{2}(x_2-x_1)[-(\psi_5^0+\phi_5^0+\xi_5^0)+(\xi_5^-+\phi_5^--\psi_5^0)x_3+(\xi_5^++\phi_5^++\psi_5^0)(1-2x_3)],\\
		P_2(x_i)=&\frac{3}{2}(x_2-x_1)[(\psi_5^0+\phi_5^0-\xi_5^0)+(\xi_5^--\phi_5^-+\psi_5^0)x_3+(\xi_5^+-\phi_5^+-\psi_5^0)(1-2x_3)].
		\end{align}}
	\item Twist-6 LCDAs\\
	{\footnotesize
		\begin{align}
		V_6(x_i)=&2[\phi_6^0+\phi_6^+(1-3x_3)],\\
		A_6(x_i)=&2(x_2-x_1)\phi_6^-,\\
		T_6(x_i)=&2[\phi_6^0+\frac{1}{2}(\phi_6^--\phi_6^+)(1-3x_3)],
		\end{align}}
\end{itemize}
with the values of the involved parameters being given in Table~\ref{table: parameter proton DAs}.

\begin{table}[tbh]
	\centering
	\caption{Twist classification of the proton LCDAs in Eq.~(\ref{eq:proton DAs}).}
	\begin{tabular*}{165mm}{c@{\extracolsep{\fill}}cccc}
		\toprule[1pt]
		\toprule[0.7pt]
		&twist-3&twist-4&twist-5&twist-6\\
		\toprule[0.7pt]
		Vector&$V_1$&$V_2,V_3$&$V_4,V_5$&$V_6$\\
		Pseudo-Vector&$A_1$&$A_2,A_3$&$A_4,A_5$&$A_6$\\
		Tensor&$T_1$&$T_2,T_3,T_7$&$T_4,T_5,T_8$&$T_6$\\
		Scalar&&$S_1$&$S_2$&\\
		Pesudoscalar&&$P_1$&$P_2$&\\
		\toprule[0.7pt]
		\toprule[1pt]
	\end{tabular*}\label{table: proton twist classification}
\end{table}

\begin{table*}[htbp]
	\centering
	\caption{Parameters in the proton LCDAs in units of $10^{-2}$ GeV$^2$~\cite{Braun:2000kw}. The accuracy of those parameters without  uncertainties is of order of $50\%$.}
	{\footnotesize
		\begin{tabular*}{165mm}{c@{\extracolsep{\fill}}ccccccccc}
			\toprule[1pt]
			\toprule[0.7pt]
			&$\phi_i^0$&$\phi_i^-$&$\phi_i^+$&$\psi_i^0$&$\psi_i^-$&$\psi_i^+$&$\xi_i^0$&$\xi_i^-$&$\xi_i^+$\\
			\toprule[0.7pt]
			twist-3 ($i=3$)&\hspace{0.3cm}$0.53\pm 0.05$&$2.11$&$0.57$&&&&&&\\
			\toprule[0.7pt]
			twist-4 ($i=4$)&\hspace{0.3cm}$-1.08\pm 0.47$&$3.22$&$2.12$&$1.61\pm 0.47$&$-6.13$&$0.99$&$0.85\pm 0.31$&$2.79$&$0.56$\\
			\toprule[0.7pt]
			twist-5 ($i=5$)&\hspace{0.3cm}$-1.08\pm 0.47$&$-2.01$&$1.42$&$1.61\pm .047$&$-0.98$&$-0.99$&$0.85\pm 0.31$&$-0.95$&$0.46$\\
			\toprule[0.7pt]
			twist-6 ($i=6$)&\hspace{0.3cm}$0.53\pm 0.05$&$3.09$&$-0.25$&&&&&&\\
			\toprule[0.7pt]
			\toprule[1pt]
	\end{tabular*}}\label{table: parameter proton DAs}
\end{table*}

\section{Numerical results}\label{sec:results}

\subsection{$\Lambda_b \to p$ form factors at $q^2=0$}
We present the numerical results of the $\Lambda_b \to p$ transition form factors in this subsection, which include the contributions from the $\Lambda_b$ baryon LCDAs up to twist 4 and the proton LCDAs up to twist 6. Because the PQCD predictions are more reliable in the large recoil (small $q^2$) region, we evaluate the form factors at $q^2=0$ and then extrapolate them to the whole kinematic range $0\leq q^2\leq (m_{\Lambda_b}-m_p)^2$ in order to estimate the branching ratios of the semileptonic decays $\Lambda_b \to p\ell\nu_\ell$.

We first compare in Table~\ref{table:simplified} the contributions to the form factors from the proton LCDAs of different twists by convoluting them with the hard kernels and the leading-twist $\Lambda_b$ baryon LCDA in Eq.~(\ref{simmod}). The values denoted by $\sim 0$ are smaller than $1\times 10^{-6}$, and the entries in the last column sum up the contributions from all the proton LCDAs. The result of the form factor $f_1(0)=1.9 \times 10^{-3}$ from the leading-twist  $\Lambda_b$ baryon and proton LCDAs is consistent with the previous PQCD calculations~\cite{Shih:1998pb,Lu:2009cm}. It is found that the contributions from the twist-4 proton LCDAs are much larger than from the leading-twist ones, which dominate the form factors $f_1(0)$ and $g_1(0)$. The twist-5 contributions to the other four form factors are also sizeable. Table~\ref{table:simplified} indicates clearly that the higher-twist contributions significantly enhance the $\Lambda_b\to p$ form factors. The enhancement of $B$ meson transition form factors by the higher-twist pion LCDAs was also observed~\cite{Kurimoto:2001zj}, but not as strong as in the baryonic case. 

\begin{table}[htbp]
	\centering
	\caption{Form factors in units of $10^{-3}$ from the leading-twist $\Lambda_b$ baryon LCDA in Eq.~(\ref{simmod}) and the proton LCDAs of various twists.}
	{\footnotesize
		\begin{tabular*}{158mm}{c@{\extracolsep{\fill}}ccccc}
			\toprule[1pt]
			\toprule[0.7pt]
			&Twist-3&Twist-4&Twist-5&Twist-6&Total\\
			\toprule[0.7pt]
			$f_1$&$1.9$&$6.3$&$1.0$&$-0.015$&$9.2$\\
			$f_2$&$0.12$&$-0.45$&$-0.63$&$\sim 0$&$-0.96$\\
			$f_3$&$-0.015$&$0.84$&$0.66$&$\sim 0$&$1.5$\\
			\toprule[0.7pt]
			$g_1$&$2.5$&$8.4$&$0.71$&$-0.008$&$11.6$\\
			$g_2$&$0.12$&$-0.30$&$-0.66$&$\sim 0$&$-0.84$\\
			$g_3$&$-0.027$&$0.90$&$0.64$&$\sim 0$&$1.5$\\
			\toprule[0.7pt]
			\toprule[1pt]
	\end{tabular*}}
	\label{table:simplified}
\end{table}

We present the contributions to the form factor $f_1(0)$ from various twist combinations of the $\Lambda_b$ baryon and proton LCDAs in Table~\ref{table:form factors f_1}. All the four models of the $\Lambda_b$ baryon LCDAs are covered. The two theoretical uncertainties in the total results are estimated from the variations of the parameters in the $\Lambda_b$ baryon LCDAs introduced in Sec.~\ref{sec:LbDAs}, and in the proton LCDAs listed in Table~\ref{table: parameter proton DAs}, respectively. It is noticed that the dominant contribution to $f_1(0)$ comes from the combination of the twist-4 $\Lambda_b$ baryon LCDA and the twist-5 proton LCDAs, and the leading-twist contribution is about two orders of magnitude lower than the dominant one, no matter which model of the $\Lambda_b$ baryon LCDAs is employed. It further confirms that the higher-twist contributions are crucial for the $\Lambda_b\to p$ form factors at the scale of the $b$ quark mass. We point out that the twist-6 proton LCDAs, as combined with the first three models of the $\Lambda_b$ baryon LCDAs, do give contributions smaller than the leading-twist ones.
Note that the $\Lambda_b$ baryon LCDA in Eqs.~(\ref{eq:simplified}) and (\ref{simmod}), despite of being classified as twist 2, contains some twist-3 components through the $m_{\Lambda_b}$ term actually (but without the mixture from the twist-4 component). Thus, the results in Table~\ref{table:simplified} should be compared to those without the twist-4 $\Lambda_b$ baryon LCDA in Table~\ref{table:form factors f_1}. It is then reasonable that the contribution from the twist-4 proton LCDAs is more important than the one from the twist-5 proton LCDAs in Table~\ref{table:simplified}.

\begin{table}[htbp]
	\footnotesize
	\centering
	\caption{Form factor $f_1(0)$ from various twist combinations of the $\Lambda_b$ baryon and proton LCDAs. The theoretical uncertainties of the total results are attributed to the variations of the relevant parameters in the $\Lambda_b$ baryon LCDAs and in the proton LCDAs, respectively.}
	\begin{tabular*}{158mm}{c@{\extracolsep{\fill}}ccccc}
		\toprule[1pt]
		\toprule[1pt]
		&twist-3&twist-4&twist-5&twist-6&total\\
		\toprule[1pt]
		exponential&&&&&\\
		twist-2&0.0007&-0.00007&-0.0005&-0.000003&0.0001\\
		twist-$3^{+-}$&-0.0001&0.002&0.0004&-0.000004&0.002\\
		twist-$3^{-+}$&-0.0002&0.0060&0.000004&0.00007&0.006\\
		twist-4&0.01&0.00009&0.25&0.0000007&0.26\\
		total&0.01&0.008&0.25&0.00007&$0.27\pm0.09\pm0.07$\\
		\toprule[0.7pt]
		free parton &&&&&\\
		twist-2&0.0006&-0.00007&-0.0005&-0.000002&0.0001\\
		twist-$3^{+-}$&-0.0001&0.002&0.0003&-0.00001&0.002\\
		twist-$3^{-+}$&-0.0002&0.006&0.00003&0.00005&0.005\\
		twist-4&0.009&0.0005&0.22&$\sim 0$&0.23\\
		total&0.009&0.008&0.22&0.00004&$0.24\pm0.07\pm0.06$\\
		\toprule[0.7pt]
		Gegenbauer-1&&&&&\\
		twist-2&0.075&-0.003&-0.063&-0.0004&0.009\\
		twist-$3^{+-}$&-0.008&0.17&0.035&0.0003&0.19\\
		twist-$3^{-+}$&-0.015&0.45&0.001&0.008&0.45\\
		twist-4&0.92&0.01&2.32&0.0002&2.41\\
		total&0.97&0.63&2.32&0.008&2.48\\
		\toprule[0.7pt]
		Gegenbauer-2&&&&&\\
		twist-2&0.00006&0.000003&-0.00002&$\sim 0$&0.00005\\
		twist-$3^{+-}$&$\sim 0$&0.006&-0.0003&-0.0004&0.005\\
		twist-$3^{-+}$&0.00002&-0.002&-0.00004&-0.0005&-0.003\\
		twist-4&0.014&0.001&0.35&$\sim 0$&0.36\\
		total&0.014&0.006&0.34&-0.0008&0.37\\
		\toprule[1pt]
		\toprule[1pt]
	\end{tabular*}
	\label{table:form factors f_1}
\end{table}

\begin{table}[htbp]
	\scriptsize
	\centering
	\caption{Numerators $h_{ij}$ of the integrands in the factorization formula for the form factor $f_1(0)$ from the diagram $D_7$ in Fig.~\ref{fig:feynman} with $i=2,3^{+-},3^{-+},4$ and $j=3,4,5,6$. An overall coefficient $\mathcal{C}m_{\Lambda_b}^3f_{\Lambda_b} /(64\sqrt{2}N_c^2)$ with the color factor $\mathcal{C}=8/3$ is implicit for the entries.}
	\begin{tabular}{ccc}
		\toprule[1pt]
		\toprule[0.7pt]
		$h_{ij}$&twist-3&twist-4\\
		\toprule[0.7pt]
		twist-2&$0$&$r\psi_24(x_1-1)x_3(-V_2+V_3+A_2+A_3+T_3+T_7+S_1-P_1)$\\
		twist-$3^{+-}$&$\psi_3^{+-}2x_3(1-x_1)(V_1+A_1)$&$r\psi_3^{+-}2x_3(V_3-A_3)$\\
		twist-$3^{-+}$&$0$&$r\psi_3^{-+}2x_3(2T_2+T_3-T_7+S_1+P_1)$\\
		twist-4&$\psi_48x_3(-T_1)$&$r\psi_44(x_1-1)(1-x_2^\prime)(V_2-V_3-A_2-A_3)$\\
		\toprule[0.7pt]
		\toprule[0.7pt]
		$h_{ij}$&twist-5&twist-6\\
		\toprule[0.7pt]
		twist-2&$r^2\psi_24x_3(-V_4+V_5-A_4-A-5)$&$r^3\psi_28(1-x_1)(1-x_2^\prime)T_6$\\
		twist-$3^{+-}$&$r^2\psi_3^{+-}2(x_1-1)(1-x_2^\prime)(T_4+2T_5-T_8+S_2+P_2)$&$0$\\
		twist-$3^{-+}$&$r^2\psi_3^{-+}2(x_1-1)(1-x_2^\prime)(V_4-A_4-T_8)$&$r^3\psi_3^{-+}2(1-x_2^\prime)(-V_6-A_6)$\\
		twist-4&$r^2\psi_44(1-x_2^\prime)(V_4-V_5+A_4+A_5+T_4+T_8+S_2-P_2)$&$0$\\
		\toprule[0.7pt]
		\toprule[1pt]
	\end{tabular}\label{table:hardkernelfunction}
\end{table}

\begin{figure}[tbp]
	\centering
	\includegraphics[width=6in]{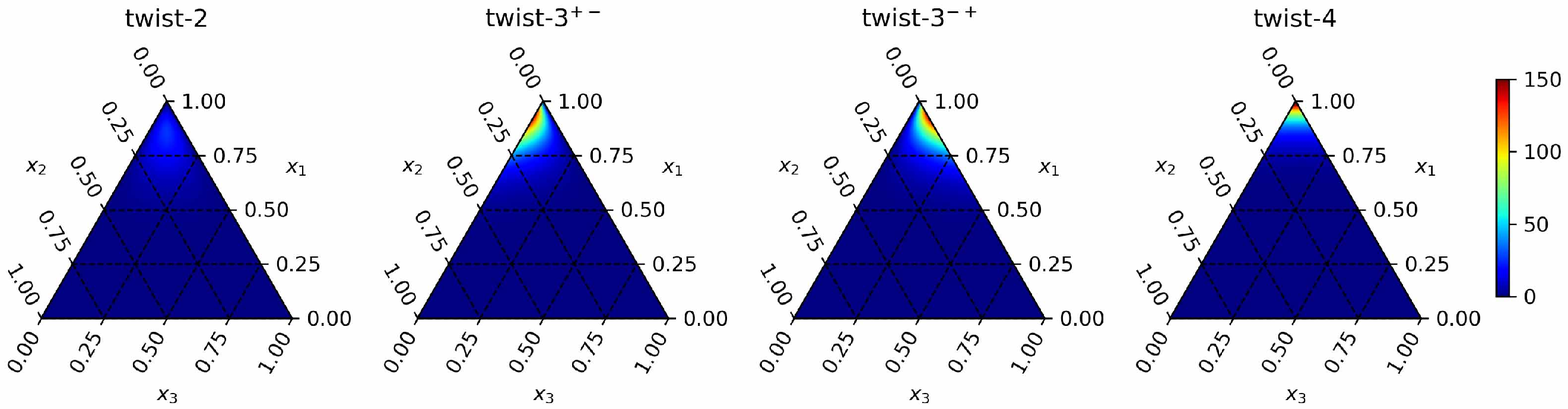}
	\includegraphics[width=6in]{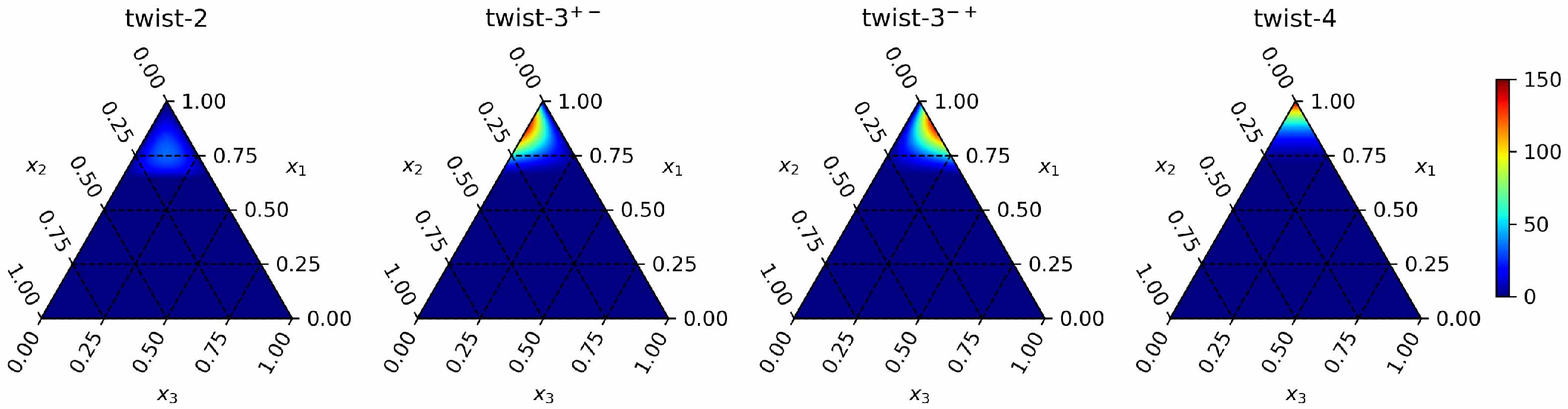}
	\caption{Dependencies of the $\Lambda_b$ baryon LCDAs $\psi_2$, $\psi_3^{+-}$, $\psi_3^{-+}$ and $\psi_4$ on the momentum fractions $x_i$ for the exponential model (Top) and the free-parton model (Bottom) proposed in \cite{Bell:2013tfa}. Each point inside the triangles satisfies the relation $x_1+x_2+x_3=1$.}
	\label{fig:EXPmodel}
\end{figure}

It is seen in Table~\ref{table:form factors f_1} that the values of $f_1(0)$ derived from the two Gegenbauer models for the $\Lambda_b$ baryon LCDAs are quite different, and those from the exponential and free-parton models are close to each other. Hence, we will focus on the latter for the numerical analysis and discussion hereafter. To understand why the power-suppressed contribution incredibly surpasses the leading-power one, we take a closer look at the behaviors of the integrands in the factorization formula for $f_1(0)$. We exhibit in Table~\ref{table:hardkernelfunction} the numerators $h_{ij}$ of the integrands from the diagram $D_7$ in Fig.~\ref{fig:feynman}, which are proportional to the products of the $\Lambda_b$ baryon LCDAs of twists $i=2,3^{+-},3^{-+},4$ and the proton LCDAs of twists $j=3,4,5,6$. The powers in the small mass ratio $r=m_p/m_{\Lambda_b}$ manifest the $1/m_{b}$ suppression associated with the higher-twist proton LCDAs. Comparing $h_{45}$ with $h_{44}$, one finds that the former contains one more power of $r$, but the latter acquires an additional factor $1-x_1$. We then illustrate the dependencies of the $\Lambda_b$ baryon LCDAs of different twists on the three momentum fractions $x_{1,2,3}$ in Fig.~\ref{fig:EXPmodel} for the exponential and free-parton models. It is obvious that these two models show similar behaviors with the twist-4 LCDAs strongly peaking around $x_1\approx 1$, where the $b$ quark carries most of the $\Lambda_b$ baryon momentum. It turns out that the factor $1-x_1$ yields a severe suppression on $h_{44}$, and $h_{45}$ contributes dominantly, although it is down by a power of $1/m_b$.

\begin{figure}[tbp]
	\centering
	\includegraphics[scale=0.2]{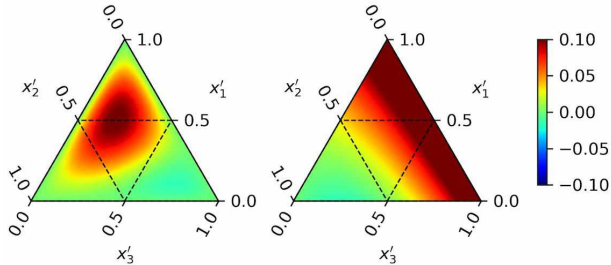}
	\caption{Dependencies of the leading-twist proton LCDAs $V_1-A_1+2T_1$ in Eq.~(\ref{eq:h23}) (Left) and of the twist-5 proton LCDAs $V_4-V_5+A_4+A_5+T_4+T_8+S_2-P_2$ in Eq.~(\ref{eq:h45}) (Right) on the momentum fractions $x_i^\prime$.}
	\label{fig:compareProton}
\end{figure}


The leading-twist LCDAs contribute to the $\Lambda_b\to p$ form factors only through the diagrams $D_1$ and $D_2$, but not through the diagram $D_7$, at the level of the theoretical accuracy in the present work. We thus compare $h_{45}$ with the leading-twist one $h_{23}$,
\begin{align}
h_{23}({\rm from}~ D_1)&={\mathcal{C}m_{\Lambda_b}^3\over 64\sqrt{2}N_c^2}f_{\Lambda_b} \psi_2 4(1-x_2)(V_1-A_1+2T_1),\label{eq:h23}\\
h_{45}({\rm from}~ D_7)&={\mathcal{C}m_{\Lambda_b}^3\over 64\sqrt{2}N_c^2}f_{\Lambda_b}r^2 \psi_4(1-x_2^\prime)(V_4-V_5+A_4+A_5+T_4+T_8+S_2-P_2),\label{eq:h45}
\end{align}
with the color factor $\mathcal{C}=8/3$. As indicated in Fig.~\ref{fig:EXPmodel}, $\psi_4$ in the
exponential and free-parton models exhibit the maxima near the endpoint $x_1\approx 1$, i.e., $(x_2+x_3)\approx 0$, and decrease with $x_2+x_3$ owing to the factors $e^{-(x_2+x_3)m_{\Lambda_b}/\omega_0}$ and $(2\bar{\Lambda}-x_2m_{\Lambda_b}-x_3m_{\Lambda_b})^3$, respectively. However, $\psi_2$, being proportional to $x_2x_3$, diminishes around the endpoint region.
Figure~\ref{fig:compareProton} shows that the combination of the higher-twist LCDAs $V_4-V_5+A_4+A_5+T_4+T_8+S_2-P_2$ is also larger than the leading-twist one $V_1-A_1+2T_1$ in the endpoint region $x_1^\prime\approx 1$, i.e., $x_2^\prime\approx 0$. The virtual particle propagators in the diagram $D_7$, being proportional to $1/(1-x_1)$ and $1/x_2^\prime$ roughly (see the factorization formulas in Appendix~\ref{app:amplitude}), induce further enhancement as $x_1\approx 1$ and $x_2^\prime\approx 0$. The above endpoint behaviors explain why the higher-twist LCDAs of both baryons overcome the power suppression from $1/m_b$ and remarkably increase the contributions to the form factors.

\begin{figure}[htbp]
	\centering
	\includegraphics[scale=0.5]{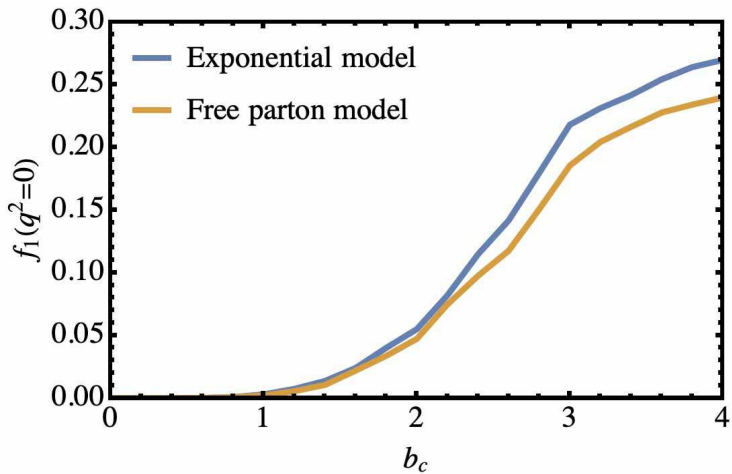}
	\caption{Dependence of the form factor $f_1(0)$ on the cutoff $b_c$.}
	\label{fig:bcEachSide}
\end{figure}
\begin{figure}[htbp]
	\centering
	\includegraphics[scale=0.7]{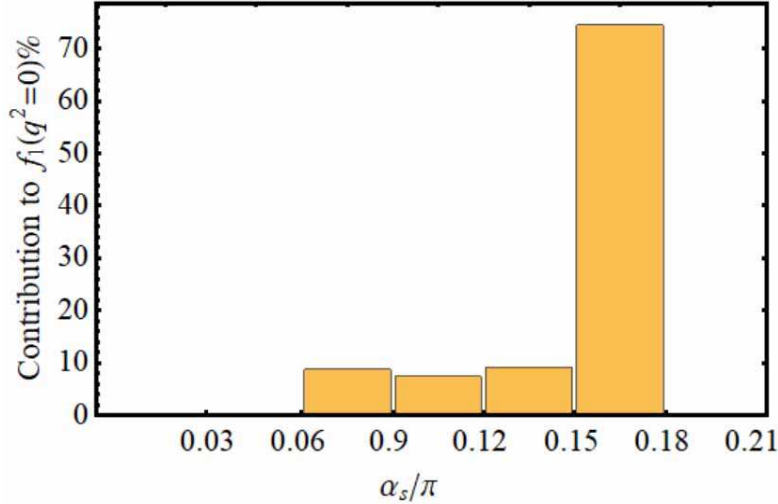}
	\caption{Contributions to the $\Lambda_b\to p$ transition form factor $f_1(q^2=0)$ from different ranges of $\alpha_s/\pi$ for the exponential model.}
	\label{fig:alphasBin}
\end{figure}

The huge enhancement caused by the higher-twist baryon LCDAs warns us to investigate how the contribution to the form factor $f_1(q^2=0)$ is distributed in the impact-parameter space. We truncate the integrations over $b_i^\prime$ on the proton side at a common upper bound $b_c<1/\Lambda_{\rm QCD}$, as done in Ref.~\cite{Li:1992nu}. The curves for both the exponential and free-parton models increase from $0$ and become flat gradually as $b_c\to 1/\Lambda_{\rm QCD}$ in Fig.~\ref{fig:bcEachSide}, implying that the Sudakov suppression on the long-distance contribution is effective enough. Since the hard $t$ is chosen as the maximum of all the scales involved in the $\Lambda_b\to p$ transition, Fig.~\ref{fig:bcEachSide} confirms that the increase of $f_1(0)$ is not attributed to nonperturbative dynamics from $b_c\sim 1/\Lambda_{\rm QCD}$. 
In addition, the contributions to the $\Lambda_b\to p$ transition form factor $f_1(0)$ from different ranges of $\alpha_s/\pi$ for the exponential model are displayed in Fig.~\ref{fig:alphasBin}. It implies that most of the contribution comes from the range with $\alpha_s/\pi < 0.2$, and that the contribution to a heavy-to-light baryonic transition form factor is indeed perturbative in the PQCD approach~\cite{Lu:2002ny}.

\begin{table}[tbp]
	\centering
	\caption{Form factors $f_1(0)$, $f_2(0)$, $g_1(0)$ and $g_2(0)$ from the exponential and free-parton models of the $\Lambda_b$ baryon LCDAs obtained in this work and in the NRQM, LCSR, 3-point QSR, lattice and previous PQCD studies. The theoretical uncertainties of our results in this work are attributed to the variations of the relevant parameters in the $\Lambda_b$ baryon LCDAs and in the proton LCDAs, respectively. The individual errors have been added in quadrature.}
	\begin{tabular*}{158mm}{c@{\extracolsep{\fill}}cccc}
		\toprule[1pt]
		\toprule[0.7pt]
		&$f_1(0)$&$f_2(0)$&$g_1(0)$&$g_2(0)$\\
		\toprule[0.7pt]
		NRQM~\cite{Mohanta:2000nk}&0.043&&&\\
		heavy-LCSR~\cite{Wang:2009hra}&$0.023^{+0.006}_{-0.005}$&&$0.023^{+0.006}_{-0.005}$&\\
		light-LCSR-$\mathcal{A}$~\cite{Khodjamirian:2011jp}&$0.14^{+0.03}_{-0.03}$&$-0.054^{+0.016}_{-0.013}$&$0.14^{+0.03}_{-0.03}$&$-0.028^{+0.012}_{-0.009}$\\
		light-LCSR-$\mathcal{P}$~\cite{Khodjamirian:2011jp}&$0.12^{+0.03}_{-0.04}$&$-0.047^{+0.015}_{-0.013}$&$0.12^{+0.03}_{-0.03}$&$-0.016^{+0.007}_{-0.005}$\\
		QCD-light-LCSR~\cite{Huang:2004vf}&0.018&-0.028&0.018&-0.028\\
		HQET-light-LCSR~\cite{Huang:2004vf}&-0.002&-0.015&&\\
		relativistic quark model~\cite{Faustov:2016pal}&0.169&0.009&0.196&-0.00004\\
		3-point QSR~\cite{Huang:1998rq}&0.22&0.0071&&\\
		lattice~\cite{Detmold:2015aaa}&$0.22\pm0.08$&$0.04\pm0.12$&$0.12\pm0.14$&$0.04\pm0.31$\\
		PQCD~\cite{Lu:2009cm}&$2.2^{+0.8}_{-0.5}\times10^{-3}$&&&\\
		\toprule[0.7pt]
		this work (exponential)&$0.27\pm0.12$&$0.008\pm0.005$&$0.31\pm 0.16$&$0.014\pm0.008$\\
		this work (free parton)&$0.24\pm0.10$&$0.007\pm 0.004$&$0.27\pm0.13$&$0.014\pm0.010$\\
		\toprule[0.7pt]
		\toprule[1pt]
	\end{tabular*}\label{table:literatures}
\end{table}

We compare our predictions for the form factors $f_1(0)$, $f_2(0)$, $g_1(0)$ and $g_2(0)$ in Table~\ref{table:literatures} with those from the nonrelativistic quark model (NRQM)~\cite{Mohanta:2000nk}, LCSR~\cite{Wang:2009hra,Khodjamirian:2011jp,Huang:2004vf}, relativistic quark model~\cite{Faustov:2016pal}, 3-point QSR~\cite{Huang:1998rq}, lattice QCD~\cite{Detmold:2015aaa} and previous PQCD studies~\cite{Lu:2009cm}. The consistency with those from light-LCSR-$\mathcal{A}$, light-LCSR-$\mathcal{P}$ and lattice QCD hints that the PQCD approach is an effective framework for analyzing exclusive heavy baryon decays, once the higher-twist contributions are taken into account. 


The other five form factors $f_{2,3}(0)$ and $g_{1,2,3}(0)$ are also computed using the exponential and free-parton models for the $\Lambda_b$ baryon LCDAs, and the outcomes are gathered in Tables~\ref{table:form factors f_2}-\ref{table:form factors g_3}. Similarly, these form factors receive large contributions from the higher-twist LCDAs, and follow the relations postulated in HQET, $f_1\approx g_1$ and $f_2\approx f_3\approx g_2\approx g_3\approx 0$~\cite{Manohar:2000dt,Mannel:1990vg}.

\begin{table}[htbp]
	\small
	\centering
	\caption{The same as Table~\ref{table:form factors f_1} but for the form factor $f_2(0)$ in unit of $10^{-3}$.}
	\begin{tabular*}{158mm}{c@{\extracolsep{\fill}}ccccc}
		\toprule[1pt]
		\toprule[1pt]
		&twist-3&twist-4&twist-5&twist-6&total\\
		\toprule[1pt]
		exponential &&&&&\\
		twist-2&0.021&0.003&-0.021&0.004&0.007\\
		twist-$3^{+-}$&-0.014&0.079&-0.12&$\sim 0$&-0.053\\
		twist-$3^{-+}$&-0.048&0.19&-0.041&0.002&0.11\\
		twist-4&0.28&-0.86&8.04&-0.017&7.45\\
		total&0.24&-0.58&7.86&-0.011&$7.5\pm3.9\pm2.7$\\
		\toprule[0.7pt]
		free parton &&&&&\\
		twist-2&0.018&0.007&-0.018&0.004&0.011\\
		twist-$3^{+-}$&-0.008&0.076&-0.11&$\sim 0$&-0.04\\
		twist-$3^{-+}$&-0.033&0.18&-0.033&0.002&0.12\\
		twist-4&0.26&-0.60&7.11&-0.017&6.75\\
		total&0.24&-0.34&6.95&-0.012&$6.8\pm3.2\pm2.9$\\
		\toprule[1pt]
		\toprule[1pt]
	\end{tabular*}
	\label{table:form factors f_2}
\end{table}

\begin{table}[htbp]
	\small
	\centering
	\caption{The same as Table~\ref{table:form factors f_1} but for the form factor $f_3(0)$ in unit of $10^{-3}$.}
	\begin{tabular*}{158mm}{c@{\extracolsep{\fill}}ccccc}
		\toprule[1pt]
		\toprule[1pt]
		&twist-3&twist-4&twist-5&twist-6&total\\
		\toprule[1pt]
		exponential &&&&&\\
		twist-2&-0.022&0.020&0.013&-0.004&0.007\\
		twist-$3^{+-}$&0.046&-0.047&0.001&$\sim 0$&0.001\\
		twist-$3^{-+}$&0.11&-0.19&0.038&-0.002&-0.037\\
		twist-4&-0.31&0.77&-7.28&0.022&-6.8\\
		total&-0.17&0.55&-7.22&0.016&$-6.8\pm4.7\pm4.1$\\
		\toprule[0.7pt]
		free parton &&&&&\\
		twist-2&-0.020&0.015&0.011&-0.004&0.002\\
		twist-$3^{+-}$&0.041&-0.041&0.006&$\approx 0$&0.007\\
		twist-$3^{-+}$&0.097&-0.18&0.032&-0.002&-0.05\\
		twist-4&-0.29&0.51&-6.38&-0.022&-6.1\\
		total&-0.17&0.30&-6.33&0.017&$-6.2\pm4.3\pm3.8$\\
		\toprule[1pt]
		\toprule[1pt]
	\end{tabular*}
	\label{table:form factors f_3}
\end{table}

\begin{table}[htbp]
	\small
	\centering
	\caption{The same as Table~\ref{table:form factors f_1} but for the form factor $g_1(0)$.}
	\begin{tabular*}{158mm}{c@{\extracolsep{\fill}}ccccc}
		\toprule[1pt]
		\toprule[1pt]
		&twist-3&twist-4&twist-5&twist-6&total\\
		\toprule[1pt]
		exponential &&&&&\\
		twist-2&0.0008&-0.00003&-0.0006&0.00002&0.0002\\
		twist-$3^{+-}$&-0.0001&0.003&-0.0003&-0.000006&0.002\\
		twist-$3^{-+}$&-0.0004&0.007&-0.0002&0.00008&0.006\\
		twist-4&0.011&-0.004&0.29&-0.0001&0.30\\
		total&0.011&0.006&0.29&-0.00001&$0.31\pm0.13\pm0.10$\\
		\toprule[0.7pt]
		free parton &&&&&\\
		twist-2&0.0007&-0.00003&-0.0005&0.00002&0.0002\\
		twist-$3^{+-}$&-0.0001&0.002&-0.0003&-0.00001&0.002\\
		twist-$3^{-+}$&-0.0003&0.007&-0.0002&0.00007&0.006\\
		twist-4&0.010&-0.00&0.25&-0.00011&0.26\\
		total&0.010&0.007&0.25&-0.00004&$0.27\pm0.11\pm0.09$\\
		\toprule[1pt]
		\toprule[1pt]
	\end{tabular*}
	\label{table:form factors g_1}
\end{table}

\begin{table}[htbp]
	\small
	\centering
	\caption{The same as Table~\ref{table:form factors f_1} but for the form factor $g_2(0)$ in unit of $10^{-3}$.}
	\begin{tabular*}{158mm}{c@{\extracolsep{\fill}}ccccc}
		\toprule[1pt]
		\toprule[1pt]
		&twist-3&twist-4&twist-5&twist-6&total\\
		\toprule[1pt]
		exponential &&&&&\\
		twist-2&0.043&-0.014&-0.034&0.008&0.004\\
		twist-$3^{+-}$&-0.061&0.079&-0.12&$\sim 0$&-0.10\\
		twist-$3^{-+}$&-0.16&0.37&-0.082&0.005&0.13\\
		twist-4&0.58&-1.62&15.5&-0.04&14.4\\
		total&0.39&-1.18&15.3&-0.028&$14.4\pm6.6\pm4.5$\\
		\toprule[0.7pt]
		free parton &&&&&\\
		twist-2&0.039&-0.010&-0.029&0.008&0.008\\
		twist-$3^{+-}$&-0.049&0.14&-0.12&$\sim 0$&-0.031\\
		twist-$3^{-+}$&-0.13&0.35&-0.067&0.003&0.16\\
		twist-4&0.53&-1.08&13.5&-0.041&12.9\\
		total&0.39&-0.60&13.3&-0.03&$13.1\pm8.4\pm4.7$\\
		\toprule[1pt]
		\toprule[1pt]
	\end{tabular*}
	\label{table:form factors g_2}
\end{table}

\begin{table}[htbp]
	\small
	\centering
	\caption{The same as Table~\ref{table:form factors f_1} but for form factor $g_3(0)$ in unit of $10^{-3}$.}
	\begin{tabular*}{158mm}{c@{\extracolsep{\fill}}ccccc}
		\toprule[1pt]
		\toprule[1pt]
		&twist-3&twist-4&twist-5&twist-6&total\\
		\toprule[1pt]
		exponential &&&&&\\
		twist-2&$\sim 0$&0.002&$\sim 0$&$\sim 0$&0.002\\
		twist-$3^{+-}$&-0.001&-0.025&$\sim 0$&$\sim 0$&-0.025\\
		twist-$3^{-+}$&$\sim 0$&0.005&-0.002&$\sim 0$&-0.004\\
		twist-4&0.003&-0.09&$\sim 0$&$\sim 0$&-0.09\\
		total&0.002&-0.11&-0.003&$\sim 0$&$-0.11\pm0.10\pm0.09$\\
		\toprule[0.7pt]
		free parton &&&&&\\
		twist-2&$\sim 0$&$\sim 0$&$\sim 0$&$\sim 0$&0.001\\
		twist-$3^{+-}$&$\sim 0$&-0.019&$\sim 0$&$\sim 0$&-0.021\\
		twist-$3^{-+}$&$\sim 0$&0.002&-0.002&$\sim 0$&$\sim 0$\\
		twist-4&0.004&-0.088&$\sim 0$&$\sim 0$&-0.083\\
		total&0.003&-0.10&-0.004&$\sim 0$&$-0.09\pm0.08\pm0.10$\\
		\toprule[1pt]
		\toprule[1pt]
	\end{tabular*}
	\label{table:form factors g_3}
\end{table}

\subsection{Semileptonic and hadronic $\Lambda_b$ baryon decays}
To calculate the branching ratios of the semileptonic decays $\Lambda_b\to p\ell\nu_\ell$ with $\ell=e,\mu,\tau$, we need the behaviors of the $\Lambda_b\to p$ transition form factors in the whole kinematic range $0\leq q^2\leq (m_{\Lambda_b}-m_p)^2$. The form factors are first evaluated at seven points of $q^2$ in the low $q^2$ region from $0$ to $m_\tau^2$, $m_\tau$ being the $\tau$ lepton mass, where the PQCD approach is trustworthy. We then derive the form factors in the large $q^2$ region via an extrapolation from the low-$q^2$ results, assuming the parametrizations of $F(\equiv f_i,g_i)$ in the $z$-series formula~\cite{Kindra:2018ayz},
\begin{equation}\label{eq:zseries}
F(q^2)=\frac{F(0)}{1-q^2/m_{\rm pole}^2}\left\{1+\sum_{k=1}^{N}a_k[z^k(q^2)-z^k(0)]\right\}.
\end{equation}
The parameter $z$ is defined as
\begin{equation}
z(q^2)=\frac{\sqrt{t_+-q^2}-\sqrt{t_+-t_0}}{\sqrt{t_+-q^2}+\sqrt{t_+-t_0}},
\end{equation}
with $t_0=t_+(1-\sqrt{1-t_-/t_+})$ and $t_\pm=(m_{\Lambda_b}\pm m_p)^2$. The values of $F(0)$ and the poles $m_{\rm pole}$ are taken from the PQCD calculation and from~\cite{Aliev:2019thw}, respectively. We truncate the expansion in Eq.~(\ref{eq:zseries}) at $N=1$ for simplicity, and fit it to the seven low-$q^2$ inputs to determine the single free parameter $a_1$. 

\begin{figure*}[tbp]
	\centering
	\includegraphics[scale=0.5]{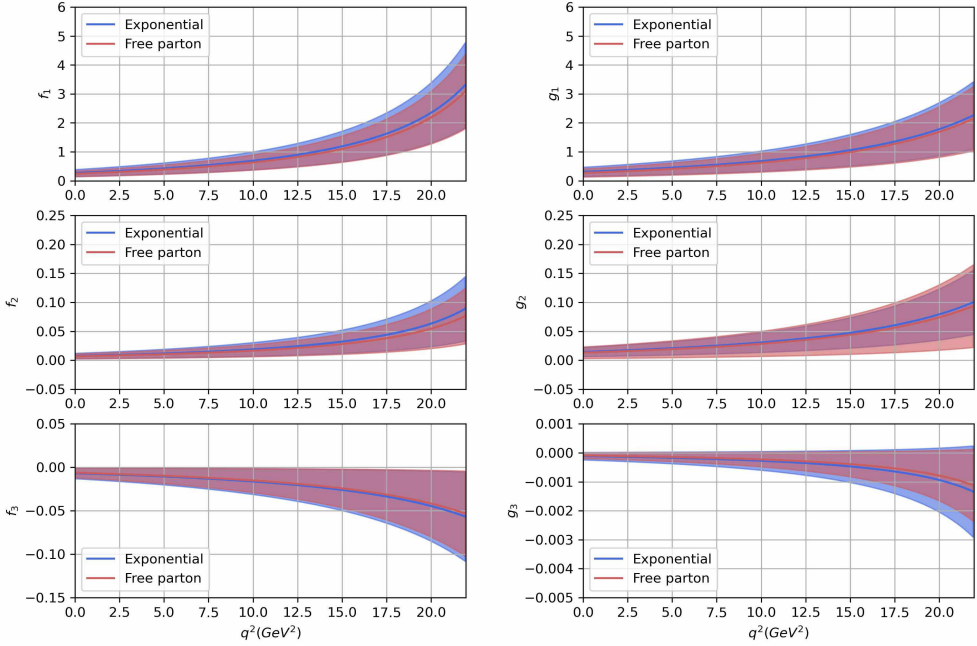}
	\caption{$q^2$ dependencies of the form factors.}
	\label{fig:formfactor}
\end{figure*}

The fit quality is measured by the  goodness of fit,
\begin{equation}
R^2=1-\frac{\sum_{j=1}^{7}(F_j-F^{\rm in}_j)^2}{\sum_{j=1}^{7}(F^{\rm in}_j-\bar{F}^{\rm in})^2},
\end{equation}
where $F^{\rm in}_j$ denote the seven PQCD inputs in the region $0\le q^2\le m_\tau^2$, $F_j$ come from Eq.~(\ref{eq:zseries}) at the same $q^2$ as $F^{\rm in}_j$, and $\bar{F}^{\rm in}$ is the mean value of the seven $F^{\rm in}_j$. We have checked that $R^2$'s in all the fits range between $0.99$ and $1$, reflecting the satisfactory quality. The resultant parameters for Eq.~(\ref{eq:zseries}) are given in Table~\ref{table:fit} with the standard errors of $a_1$ being assigned. The $q^2$ dependencies of the form factors are exhibited in Fig.~\ref{fig:formfactor}, where the bands represent the theoretical uncertainties caused by the parameters in the baryon LCDAs and by the parameter $a_1$. The absolute values of all the form factors increase with $q^2$ as expected.


\begin{table*}[tbp]
	\centering
	\caption{Values of the parameters for the form factors in Eq.~(\ref{eq:zseries}).}
	\begin{tabular*}{165mm}{c@{\extracolsep{\fill}}ccc}
		\toprule[1pt]
		\toprule[0.7pt]
		exponential&$F(0)$&$m_{\rm pole}$ (GeV)&$a_1$\\
		\toprule[0.7pt]
		$f_1$&$0.27\pm0.12$&$5.325$&$-10.6\pm2.1$\\
		$f_2$&$0.008\pm0.005$&$5.325$&$-8.5\pm0.8$\\
		$f_3$&$-0.007\pm0.006$&$5.749$&$-10.1\pm0.7$\\
		$g_1$&$0.31\pm0.16$&$5.723$&$-8.0\pm2.2$\\
		$g_2$&$0.014\pm0.008$&$5.723$&$-7.4\pm1.0$\\
		$g_3$&$-0.00011\pm0.00013$&$5.280$&$-9.1\pm0.2$\\
		\toprule[0.7pt]
		free parton&$F(0)$&$m_{\rm pole}$ (GeV)&$a_1$\\
		\toprule[0.7pt]
		$f_1$&$0.24\pm0.10$&$5.325$&$-10.8\pm2.0$\\
		$f_2$&$0.007\pm0.004$&$5.325$&$-7.7\pm1.1$\\
		$f_3$&$-0.006\pm0.006$&$5.749$&$-10.7\pm0.8$\\
		$g_1$&$0.27\pm0.14$&$5.723$&$-9.2\pm2.4$\\
		$g_2$&$0.013\pm0.010$&$5.723$&$-7.7\pm1.2$\\
		$g_3$&$-0.00009\pm0.00012$&$5.280$&$-9.4\pm0.3$\\
		\toprule[0.7pt]
		\toprule[1pt]
	\end{tabular*}\label{table:fit}
\end{table*}

\begin{figure*}[tbh]
	\centering
	\includegraphics[scale=0.7]{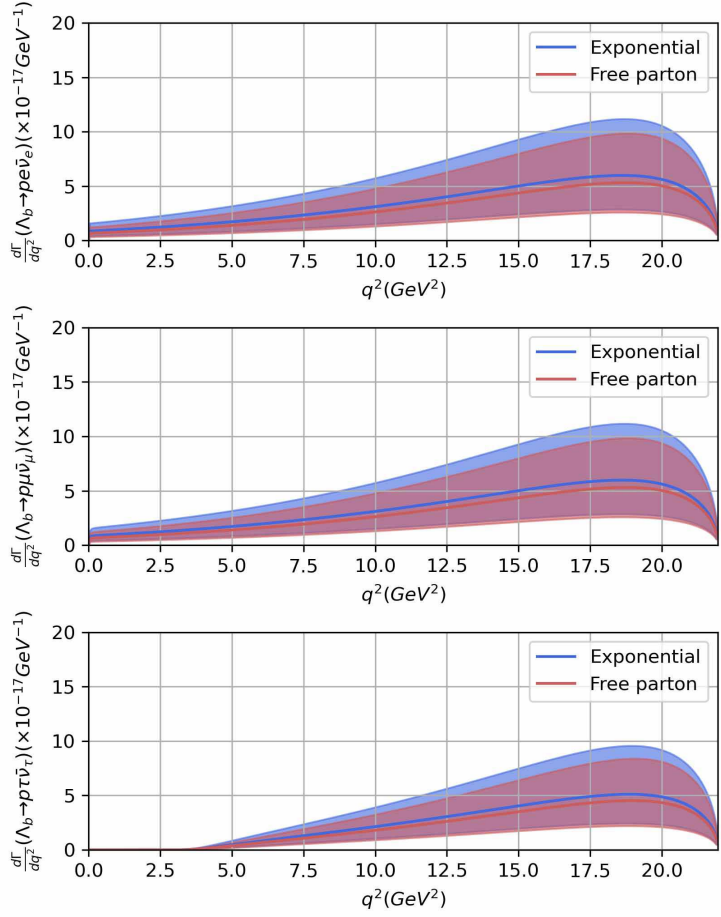}
	\caption{$q^2$ dependencies of the differential widths for the $\Lambda_b\to p \ell\bar{\nu}_\ell$ decays with $\ell=e,\nu,\tau$.}
	\label{fig:DDW}
\end{figure*}

We then predict the branching ratios of the semileptonic decays $\Lambda_b\to p\ell\bar{\nu}_\ell$ based on the form factors presented in Fig.~\ref{fig:formfactor}. We introduce the helicity amplitudes
\begin{align}
H_{\lambda_2,\lambda_W}=&H_{\lambda_2,\lambda_W}^V-H_{\lambda_2,\lambda_W}^A,\\
H_{\lambda_2,\lambda_W}^{V(A)}=&\epsilon^{\dagger\mu}(\lambda_W)\langle p,\lambda_2|V(A)|\Lambda_b,\lambda_1\rangle,
\end{align}
where $\lambda_1$, $\lambda_2$ and $\lambda_W$ denote the helicities of the $\Lambda_b$ baryon, the proton and the off-shell $W^-$ boson, respectively, $\epsilon^\mu$ is the polarization vector of the $W^-$ boson, and $V$ ($A$) labels the vector (axial vector) current. The various helicity amplitudes are then expressed, in terms of the $\Lambda_b\to p$ form factors, as~\cite{Azizi:2019tcn,Azizi:2018axf,Zhang:2019jax,Gutsche:2015mxa,Shivashankara:2015cta,Dutta:2015ueb,Zwicky:2013eda,Hiller:2021zth}
\begin{align}
H_{1/2,0}^V&=\frac{\sqrt{(m_{\Lambda_b}-m_p)^2-q^2}}{\sqrt{q^2}}\left[(m_{\Lambda_b}+m_p)f_1(q^2) + q^2f_2(q^2)\right],\nonumber\\
H_{1/2,0}^A&=\frac{\sqrt{(m_{\Lambda_b}+m_p)^2-q^2}}{\sqrt{q^2}}\left[(m_{\Lambda_b}-m_p)g_1(q^2) - q^2g_2(q^2)\right],\nonumber\\
H_{1/2,1}^V&=\sqrt{2[(m_{\Lambda_b}-m_p)^2-q^2]}\left[-f_1(q^2) - (m_{\Lambda_b}+m_p)f_2(q^2)\right],\nonumber\\
H_{1/2,1}^A&=\sqrt{2[(m_{\Lambda_b}+m_p)^2-q^2]}\left[-g_1(q^2) + (m_{\Lambda_b}-m_p)g_2(q^2)\right],\nonumber\\
H_{1/2,t}^V&=\frac{\sqrt{(m_{\Lambda_b}+m_p)^2-q^2}}{\sqrt{q^2}}\left[(m_{\Lambda_b}-m_p)f_1(q^2) + q^2f_3(q^2)\right],\nonumber\\
H_{1/2,t}^A&=\frac{\sqrt{(m_{\Lambda_b}-m_p)^2-q^2}}{\sqrt{q^2}}\left[(m_{\Lambda_b}+m_p)g_1(q^2) - q^2g_3(q^2)\right],
\end{align}
where the subscript $t$ refers to the temporal component of the helicities of the off-shell $W^-$ boson~\cite{Gutsche:2015mxa}. The above amplitudes obey the relations $H_{\lambda_2,\lambda_W}^V=H_{-\lambda_2,-\lambda_W}^V$ and $H_{\lambda_2,\lambda_W}^A=-H_{-\lambda_2,-\lambda_W}^A$.

The differential angular distributions for the semileptonic decays $\Lambda_b\to p\ell\bar{\nu}_\ell$ are given by~\cite{Li:2021qod,Azizi:2019tcn}
\begin{equation}\label{ddG}
\frac{d\Gamma(\Lambda_b\to p\ell\bar{\nu}_\ell)}{dq^2d\cos\theta_\ell}=\frac{G_F^2|V_{ub}|^2q^2|\vec{p}_p|}{512\pi^3m_{\Lambda_b}^2}\left(1-\frac{m_\ell^2}{q^2}\right)^2\left(A_1+\frac{m_\ell^2}{q^2}A_2\right),
\end{equation}
where $m_\ell$ is the charged lepton mass, $\theta_\ell$ stands for the angle between the charged lepton and the proton in the $\Lambda_b$ baryon rest frame, and the factors $A_1$, $A_2$ and $|\vec{p}_p|$ are written as
\begin{align}
A_1=&2\sin^2\theta_\ell(H_{1/2,0}^2+H_{-1/2,0}^2)+(1-\cos\theta_\ell)^2H_{1/2,1}^2+(1+\cos\theta_\ell)^2H_{-1/2,-1}^2,\\
A_2=&2\cos^2\theta_\ell(H_{1/2,0}^2+H_{-1/2,0}^2)+\sin^2\theta_\ell(H_{1/2,1}^2+H_{-1/2,-1}^2)+2(H_{1/2,t}^2+H_{-1/2,t}^2),\nonumber\\
&-4\cos\theta_\ell(H_{1/2,0}H_{1/2,t}+H_{-1/2,0}H_{-1/2,t})\\
|\vec{p}_p|=&\frac{\sqrt{m_{\Lambda_b}^4+m_p^4+q^4-2(m_{\Lambda_b}^2m_p^2+m_p^2q^2+q^2m_{\Lambda_b}^2)}}{2m_{\Lambda_b}}.
\end{align}
Integrating Eq.~(\ref{ddG}) over $\cos \theta_\ell$, we derive the differential decay widths 
\begin{equation}
\frac{d\Gamma(\Lambda_b\to p\ell\bar{\nu}_\ell)}{dq^2}=\int_{-1}^{1}\frac{d\Gamma(\Lambda_b\to p\ell\bar{\nu}_\ell)}{dq^2d\cos\theta_\ell}d\cos\theta_\ell.
\end{equation}
plotted in Fig.~\ref{fig:DDW} for $\ell=e,\mu,\tau$, which lead to the integrated branching ratios  $\mathcal{B}(\Lambda_b\to p\ell\bar \nu_\ell)=(16\pm11)\times10^{-4}$ with $\ell=e,\mu$ and $\mathcal{B}(\Lambda_b\to p\tau\bar \nu_\tau)=(11\pm7)\times10^{-4}$ for the exponential model of the $\Lambda_b$ baryon LCDAs, and $\mathcal{B}(\Lambda_b\to p\ell\bar \nu_\ell)=(14\pm10)\times10^{-4}$ with $\ell=e,\mu$ and $\mathcal{B}(\Lambda_b\to p\tau\bar \nu_\tau)=(10\pm7)\times10^{-4}$ for the free parton model. We remind that the above results for the $\Lambda_b\to p\tau\bar \nu_\tau$ decay depend on the extrapolation of the form factors to the large $q^2$ region completely. The central values of our predictions are higher than the LHCb data $\mathcal{B}(\Lambda_b\to p\mu\bar \nu_\mu)=(4.1\pm1.0)\times10^{-4}$ \cite{LHCb:2015eia} due to the larger form factors obtained in the previous subsection. However, they are still compatible with each other when the substantial theoretical uncertainties are considered. Besides, we expect that our results will decrease a bit, after the threshold Sudakov factor mentioned in the Introduction and the intrinsic impact-parameter dependencies of the baryon wave functions are included.


Next we estimate the branching ratio of the two-body hadronic decay $\Lambda_b\to p M$, with $M$ denoting a light pesudoscalar meson, in the naive factorization framework. This decay is dominated by the color-favored tree contribution, for which the naive factorization assumption is supposed to hold reasonably well. The corresponding decay amplitude is expressed as
\begin{equation}
\langle p M |\mathcal{H}_{eff}|\Lambda_b \rangle = \frac{G_F}{\sqrt{2}}V_{ub}^{\ast}V_{uq}a_1(\mu)\langle M|\bar{u}\gamma^\mu(1-\gamma_5)q|0\rangle \langle p|\bar{u}\gamma_\mu(1-\gamma_5)b|\Lambda_b\rangle,
\end{equation}
where $G_F=1.166\times 10^{-5}$ GeV$^{-2}$ is the Fermi constant, $V_{ub}$ and $V_{uq}$ are the Cabibbo-Kobayashi-Maskawa (CKM) matrix elements, and $a_1(\mu)=C_1(\mu)+C_2(\mu)/3$ represents the Wilson coefficient with $C_1(m_b)=1.076$ and $C_2(m_b)=-0.175$ at the scale of the $b$ quark mass $m_b=4.8$ GeV.

Inserting the definition of the pseudoscalar decay constant $f_M$,
\begin{equation}
\langle M(q)|\bar{u}\gamma^\mu(1-\gamma_5)d|0\rangle = -if_M q^\mu,
\end{equation}
$q$ being the pseudoscalar meson momentum, we decompose the $\Lambda_b\to p M$ decay amplitude into
\begin{equation}
{\cal M}(\Lambda_b\to pM)=i\bar{N}({\cal M}_1+{\cal M}_2\gamma_5)\Lambda_b.
\end{equation}
The functions ${\cal M}_1$ and ${\cal M}_2$ are given by
\begin{align}
{\cal M}_1=& \frac{G_F}{\sqrt{2}}V_{ub}^{\ast}V_{uq}a_1(\mu)f_M (m_{\Lambda_b}-m_p)f_1(m_M^2),\\
{\cal M}_2=& \frac{G_F}{\sqrt{2}}V_{ub}^{\ast}V_{uq}a_1(\mu)f_M(m_{\Lambda_b}+m_p)g_1(m_M^2),
\end{align}
with the pseudoscalar meson mass $m_M$. The $\Lambda_b\to p M$ decay width is then written as
\begin{equation}
\Gamma(\Lambda_b\to pM)=\frac{|\overrightarrow{p_p}|}{8\pi}\left[\frac{(m_{\Lambda_b}+m_p)^2-m_M^2}{m_{\Lambda_b}^2}|{\cal M}_1|^2+\frac{(m_{\Lambda_b}-m_p)^2-m_M^2}{m_{\Lambda_b}^2}|{\cal M}_2|^2\right],
\end{equation}
where $|\overrightarrow{p_p}|=\sqrt{[(m_{\Lambda_b}^2-(m_M+m_p)^2][m_{\Lambda_b}^2-(m_M-m_p)^2]}/(2m_{\Lambda_b})$ is the proton momentum in the rest frame of the $\Lambda_b$ baryon.

It is straightforward to get, by employing the form factors derived in the previous subsection, the $\Lambda_b\to p\pi$ branching ratio $\mathcal{B}(\Lambda_b\to p\pi^-)=(13\pm10)\times10^{-6}$ from the exponential model for the $\Lambda_b$ baryon LCDAs, and $\mathcal{B}(\Lambda_b\to p\pi^-)=(11\pm8)\times10^{-6}$ from the free-parton model, whose large theoretical uncertainties  originate from those of the form factors. These predictions are also higher than the experimental data $\mathcal{B}(\Lambda_b\to p\pi^-)=(4.5\pm0.8)\times10^{-6}$ ~\cite{ParticleDataGroup:2020ssz}, similar to the case of the semileptonic decays. 



\section{Summary}\label{sec:summary}
The heavy-to-light transition form factors are important ingredients for exclusive heavy hadron decays. The previous studies of the $\Lambda_b \to p$ form factors in the QCDF and PQCD approaches have manifested that the leading-power contribution is much smaller than indicated by the experimental data. In this paper we extended the PQCD analysis to the inclusion of the higher-twist baryon LCDAs. It was observed that the combination of the twist-4 $\Lambda_b$ baryon LCDAs and the twist-5 (twist-4) proton LCDAs dominates the contributions to the form factors $f_{1,2,3}$ and $g_{1,2}$ (the form factor $g_3$), and that the contributions from the twist-6 proton LCDAs are indeed suppressed. We have examined the distribution of the above dominant pieces in the impact-parameter space, and concluded that the enhancement of the form factors is not attributed to the long-distance dynamics. Our results for the form factors are close to those in other theoretical methods within errors, implying that the endpoint contributions from the higher-twist LCDAs can be handled appropriately in the PQCD formalism, and the framework established here is ready for systematic applications to various semileptonic and hadronic two-body decays of heavy baryons.

Based on the obtained form factors, we have estimated the branching ratios of the semileptonic decays $\Lambda_b\to p\ell\bar \nu_\ell$ and of the two-body hadronic decay $\Lambda_b \to p\pi$ under the naive factorization assumption, whose central values are higher than but still compatible with the measured ones, as the significant theoretical uncertainties are considered. It suggests that a precise knowledge of the baryon LCDAs is necessary for a rigorous comparison between theoretical predictions and experimental data. It is also urgent to derive the threshold Sudakov factor for heavy-to-light baryonic transitions, which is expected to improve the agreement with the current data by lowering our results to some extent. The intrinsic impact-parameter dependencies of the baryon wave functions can also be taken into account to achieve the same purpose. At last, the complete set of topological diagrams needs to be calculated in order to make predictions for CPV in hadroic two-body heavy baryon decays. This calculation is feasible in principle in the PQCD approach.



\section*{Acknowledgement}
We thank Yu-Ming Wang, Shan Cheng and Dong-Hao Li for useful discussions and valuable suggestions. 
This work was supported in part by the National Natural Science Foundation of China under
Grant Nos.~11775117,~11975112,~12005103,~12175218, and by the Ministry of Science and Technology of R.O.C. under Grant No. MOST-110-2811-M-001-540-MY3.
YL is also supported by the Natural Science Foundation of Jiangsu Province under
Grant No.~BK20190508 and the Research Start-up Funds of Nanjing Agricultural University.
JJH is supported by the Practice Innovation Program of Jiangsu Province under Grant No. KYCX21$\_$1320.
Y.L.S also acknowledges the Natural Science Foundation of Shandong province with Grant No. ZR2020MA093.

\appendix

\section{Factorization formulas}\label{app:amplitude}
We list below the factorization formulas for the form factor $f_1(q^2=0)$ from the Feynman diagrams $D_i$, $i=1$-14, in Fig.~\ref{fig:feynman}. Those for the other five form factors can be derived in a similar way. The last two diagrams with three-gluon vertices do not contribute due to the vanishing color factors:
{\footnotesize
	\begin{align}
	f_1^{D_1}(q^2=0)=&\mathcal{C}M_{\Lambda_b}^3 \frac{f_{\Lambda_b}}{8N_c} \frac{1}{8\sqrt{2}N_c}\int[dx]\int[dx^\prime]16\pi^2\alpha_s^2(t^{D_1})h \frac{1}{(2\pi)^5}\int b_1^\prime db_1^\prime \int b_2 db_2\nonumber\\
	&\int b_3 db_3\int d\theta_1\int d\theta_2\exp[-S^{D_1}(x,x^\prime ,b,b^\prime)]F_1(D,b_3)F_3(A,B,C,b_1^\prime,b_2),
	\end{align}}
{\footnotesize
	\begin{align}
	f_1^{D_2}(q^2=0)=&\mathcal{C}M_{\Lambda_b}^3 \frac{f_{\Lambda_b}}{8N_c} \frac{1}{8\sqrt{2}N_c}\int[dx]\int[dx^\prime]16\pi^2\alpha_s^2(t^{D_2})h \frac{1}{(2\pi)^5}\int b_1^\prime db_1^\prime \int b_2 db_2\nonumber\\
	&\int b_2^\prime db_2^\prime\int d\theta_1\int d\theta_2\exp[-S^{D_2}(x,x^\prime ,b,b^\prime)]F_1(C,b_2)F_3(A,B,D,b_1^\prime,b_3),
	\end{align}}
{\footnotesize
	\begin{align}
	f_1^{D_3}(q^2=0)=&\mathcal{C}M_{\Lambda_b}^3 \frac{f_{\Lambda_b}}{8N_c} \frac{1}{8\sqrt{2}N_c}\int[dx]\int[dx^\prime]16\pi^2\alpha_s^2(t^{D_3})h \frac{1}{(2\pi)^7}\int b_1db_1\int b_1^\prime db_1^\prime \int b_3 db_3\nonumber\\
	&\int b_3^\prime db_3^\prime\int d\theta_1\int d\theta_2\int d\theta_3\exp[-S^{D_3}(x,x^\prime ,b,b^\prime)]F_1(A,b_1+b_3-b_1^\prime-b_3^\prime)\nonumber\\
	&F_1(B,b_3+b_3^\prime)F_1(C,b_1)F_1(D,b_3^\prime),
	\end{align}}
{\footnotesize
	\begin{align}
	f_1^{D_4}(q^2=0)=&\mathcal{C}M_{\Lambda_b}^3 \frac{f_{\Lambda_b}}{8N_c} \frac{1}{8\sqrt{2}N_c}\int[dx]\int[dx^\prime]16\pi^2\alpha_s^2(t^{D_4})h \frac{1}{(2\pi)^7}\int b_1db_1\int b_1^\prime db_1^\prime \int b_3 db_3\nonumber\\
	&\int b_3^\prime db_3^\prime\int d\theta_1\int d\theta_2\int d\theta_3\exp[-S^{D_4}(x,x^\prime ,b,b^\prime)]F_1(A,b_1+b_3-b_1^\prime-b_3^\prime)\nonumber\\
	&F_1(B,b_3+b_3^\prime)F_1(C,b_3+b_3^\prime-b_1)F_1(D,b_3),
	\end{align}}
{\footnotesize
	\begin{align}
	f_1^{D_5}(q^2=0)=&\mathcal{C}M_{\Lambda_b}^3 \frac{f_{\Lambda_b}}{8N_c} \frac{1}{8\sqrt{2}N_c}\int[dx]\int[dx^\prime]16\pi^2\alpha_s^2(t^{D_5})h \frac{1}{(2\pi)^7}\int b_1db_1\int b_1^\prime db_1^\prime \int b_2 db_2\nonumber\\
	&\int b_2^\prime db_2^\prime\int d\theta_1\int d\theta_2\int d\theta_3\exp[-S^{D_5}(x,x^\prime ,b,b^\prime)]F_1(A,b_1+b_1^\prime-b_2-b_2^\prime)\nonumber\\
	&F_1(B,b_2+b_2^\prime)F_1(C,b_2+b_2^\prime-b_1)F_1(D,b_2),
	\end{align}}
{\footnotesize
	\begin{align}
	f_1^{D_6}(q^2=0)=&\mathcal{C}M_{\Lambda_b}^3 \frac{f_{\Lambda_b}}{8N_c} \frac{1}{8\sqrt{2}N_c}\int[dx]\int[dx^\prime]16\pi^2\alpha_s^2(t^{D_6})h \frac{1}{(2\pi)^7}\int b_1db_1\int b_1^\prime db_1^\prime \int b_2 db_2\nonumber\\
	&\int b_2^\prime db_2^\prime\int d\theta_1\int d\theta_2\int d\theta_3\exp[-S^{D_6}(x,x^\prime ,b,b^\prime)]F_1(A,b_1+b_1^\prime-b_2-b_2^\prime)\nonumber\\
	&F_1(B,b_2+b_2^\prime)F_1(C,b_1)F_1(D,b_2^\prime),
	\end{align}}
{\footnotesize
	\begin{align}
	f_1^{D_7}(q^2=0)=&\mathcal{C}M_{\Lambda_b}^3 \frac{f_{\Lambda_b}}{8N_c} \frac{1}{8\sqrt{2}N_c}\int[dx]\int[dx^\prime]16\pi^2\alpha_s^2(t^{D_7})h \frac{1}{(2\pi)^5}\int b_1 db_1 \int b_2^\prime db_2^\prime\nonumber\\
	&\int b_3 db_3\int d\theta_1\int d\theta_2\exp[-S^{D_7}(x,x^\prime ,b,b^\prime)]F_1(D,b_2^\prime-b_3^\prime)F_3(A,B,C,b_2^\prime,b_1),
	\end{align}}
{\footnotesize
	\begin{align}
	f_1^{D_8}(q^2=0)=&\mathcal{C}M_{\Lambda_b}^3 \frac{f_{\Lambda_b}}{8N_c} \frac{1}{8\sqrt{2}N_c}\int[dx]\int[dx^\prime]16\pi^2\alpha_s^2(t^{D_8})h \frac{1}{(2\pi)^5}\int b_1 db_1 \int b_2 db_2\nonumber\\
	&\int b_2^\prime db_2^\prime \int d\theta_1\int d\theta_2\exp[-S^{D_8}(x,x^\prime ,b,b^\prime)]F_1(C,b_2^\prime)F_3(A,B,D,b_2+b_2^\prime,b_1),
	\end{align}}
{\footnotesize
	\begin{align}
	f_1^{D_9}(q^2=0)=&\mathcal{C}M_{\Lambda_b}^3 \frac{f_{\Lambda_b}}{8N_c} \frac{1}{8\sqrt{2}N_c}\int[dx]\int[dx^\prime]16\pi^2\alpha_s^2(t^{D_9})h \frac{1}{(2\pi)^7}\int b_1db_1\int b_1^\prime db_1^\prime \int b_3 db_3\nonumber\\
	&\int b_3^\prime db_3^\prime\int d\theta_1\int d\theta_2\int d\theta_3\exp[-S^{D_9}(x,x^\prime ,b,b^\prime)]F_1(A,b_1+b_1^\prime-b_3-b_3^\prime)\nonumber\\
	&F_1(B,b_3+b_3^\prime)F_1(C,b_3+b_3^\prime-b_1^\prime)F_1(D,b_3^\prime),
	\end{align}}
{\footnotesize
	\begin{align}
	f_1^{D_{10}}(q^2=0)=&\mathcal{C}M_{\Lambda_b}^3 \frac{f_{\Lambda_b}}{8N_c} \frac{1}{8\sqrt{2}N_c}\int[dx]\int[dx^\prime]16\pi^2\alpha_s^2(t^{D_{10}})h\frac{1}{(2\pi)^7}\int b_1db_1\int b_1^\prime db_1^\prime \int b_3 db_3\nonumber\\
	&\int b_3^\prime db_3^\prime\int d\theta_1\int d\theta_2\int d\theta_3\exp[-S^{D_{10}}(x,x^\prime ,b,b^\prime)]F_1(A,b_1+b_1^\prime-b_3-b_3^\prime)\nonumber\\
	&F_1(B,b_3+b_3^\prime)F_1(C,b_1^\prime)F_1(D,b_3),
	\end{align}}
{\footnotesize
	\begin{align}
	f_1^{D_{11}}(q^2=0)=&\mathcal{C}M_{\Lambda_b}^3 \frac{f_{\Lambda_b}}{8N_c} \frac{1}{8\sqrt{2}N_c}\int[dx]\int[dx^\prime]16\pi^2\alpha_s^2(t^{D_{11}})h \frac{1}{(2\pi)^7}\int b_1db_1\int b_1^\prime db_1^\prime \int b_2 db_2\nonumber\\
	&\int b_2^\prime db_2^\prime\int d\theta_1\int d\theta_2\int d\theta_3\exp[-S^{D_{11}}(x,x^\prime ,b,b^\prime)]F_1(A,b_1+b_1^\prime-b_2-b_2^\prime)\nonumber\\
	&F_1(B,b_2+b_2^\prime)F_1(C,b_2^\prime)F_1(D,b_2+b_2^\prime-b_1^\prime),
	\end{align}}
{\footnotesize
	\begin{align}
	f_1^{D_{12}}(q^2=0)=&\mathcal{C}M_{\Lambda_b}^3 \frac{f_{\Lambda_b}}{8N_c} \frac{1}{8\sqrt{2}N_c}\int[dx]\int[dx^\prime]16\pi^2\alpha_s^2(t^{D_{12}})h \frac{1}{(2\pi)^7}\int b_1db_1\int b_1^\prime db_1^\prime \int b_2 db_2\nonumber\\
	&\int b_2^\prime db_2^\prime\int d\theta_1\int d\theta_2\int d\theta_3\exp[-S^{D_{12}}(x,x^\prime ,b,b^\prime)]F_1(A,b_1+b_1^\prime-b_2-b_2^\prime)\nonumber\\
	&F_1(B,b_2+b_2^\prime)F_1(C,b_1^\prime)F_1(D,b_2),
	\end{align}}
{\footnotesize
	\begin{align}
	f_1^{D_{13}}(q^2=0)=&\mathcal{C}M_{\Lambda_b}^3 \frac{f_{\Lambda_b}}{8N_c} \frac{1}{8\sqrt{2}N_c}\int[dx]\int[dx^\prime]16\pi^2\alpha_s^2(t^{D_{13}})h \frac{1}{(2\pi)^5}\int b_2 db_2 \int b_2^\prime db_2^\prime\nonumber\\
	&\int b_3 db_3\int d\theta_1\int d\theta_2\exp[-S^{D_{13}}(x,x^\prime ,b,b^\prime)]F_2(A,B,b_2+b_2^\prime)F_1(C,b_2^\prime)F_1(D,b_3),
	\end{align}}
{\footnotesize
	\begin{align}
	f_1^{D_{14}}(q^2=0)=&\mathcal{C}M_{\Lambda_b}^3 \frac{f_{\Lambda_b}}{8N_c} \frac{1}{8\sqrt{2}N_c}\int[dx]\int[dx^\prime]16\pi^2\alpha_s^2(t^{D_{14}})h \frac{1}{(2\pi)^5}\int b_2 db_2 \int b_2^\prime db_2^\prime\nonumber\\
	&\int b_3 db_3\int d\theta_1\int d\theta_2\exp[-S^{D_{14}}(x,x^\prime ,b,b^\prime)]F_2(A,B,b_2+b_2^\prime)F_1(C,b_2)F_1(D,b_3^\prime),
	\end{align}}
where $\mathcal{C}=8/3$ is the color factor and $[dx]\equiv dx_1dx_2dx_3\delta(1-x_1-x_2-x_3)$, $[dx']$ is defined analogously, the auxiliary functions $A,B,C$ and $D$ in Table~\ref{table:ABCD} are related to the denominators of the four propagators in each diagram, and the hard kernels $h$ are collected in Tables~\ref{table:hardfunctions1}-\ref{table:hardfunctions3}. The exponent $S^{D_i}$ is the sum of the total exponents from the $\Lambda_b$ and proton wave functions with the hard scale $t^{D_i}$ involved in the diagram $D_i$. The functions $F_1, F_2$ and $F_3$ are written, in terms of the Fourier integrals, as
{\footnotesize
	\begin{align}
	F_1(A,b)=&\int d^2k_T\frac{e^{i\textbf{k}_T\cdot\textbf{b}}}{k^2+A}=2\pi \left\{K_0(\sqrt{A}b)\theta(A)+\frac{\pi i}{2}\left[J_0(\sqrt{-A}b)+iN_0(\sqrt{-A}b)\right]\theta(-A)\right\},
	\end{align}}
{\footnotesize
	\begin{align}
	F_2(A,B,b)=&\int d^2k_T\frac{e^{i\textbf{k}_T\cdot\textbf{b}}}{(k^2+A)(k^2+B)}\nonumber\\
	=&\pi\int_{0}^{1}dz\frac{b}{\sqrt{|Z_1|}}\left\{K_1(\sqrt{Z_1}b)\theta(Z_1)
	+\frac{\pi}{2}\left[N_1(\sqrt{-Z_1}b)-iJ_1(\sqrt{-Z_1}b)\right]\theta(-Z_1)\right\},
	\end{align}}
{\footnotesize
	\begin{align}
	F_3(A,B,C,b_1,b_2)=&\int d^2k_{1T}\int d^2k_{2T}\frac{e^{i(\textbf{k}_{1T}\cdot\textbf{b}_1+\textbf{k}_{2T}\cdot\textbf{b}_2)}}{(k_1^2+A)(k_2^2+B)((k_1+k_2)^2+C)}\nonumber\\
	=&\pi^2\int_{0}^{1}\frac{dz_1dz_2}{z_1(1-z_1)}\frac{\sqrt{X_2}}{\sqrt{|Z_2|}}\nonumber\\
	&\times\left\{K_1(\sqrt{X_2Z_2})\theta(Z_2)+\frac{\pi}{2}\left[N_1(\sqrt{-X_2Z_2})-iJ_1(\sqrt{-X_2Z_2})\right]\theta(-Z_2)\right\},
	\end{align}}
in which $J_n$ ($N_n$) is the Bessel function of the first (second) kind, $K_n$ follows the relation
\begin{align}
K_n(-iz)=\frac{\pi i}{2}e^{(in\pi)/2}\left[J_n(z)+iN_n(z)\right],
\end{align}
and the variables $Z_1$, $Z_2$ and $X_2$ are given by
\begin{align}
Z_1=&Az+B(1-z),\\
Z_2=&A(1-z_2)+\frac{z_2}{z_1(1-z_1)}\left[B(1-z_1)+Cz_1\right],\\
X_2=&(b_1-z_1b_2)^2+\frac{z_1(1-z_1)}{z_2}b_2^2,
\end{align}
with Feynman parameters $z$'s. 
Because $A$, $Z_1$, $Z_2$ and $X_2$ are all positive, the imaginary parts of the above functions $F_{1,2,3}$ do not contribute.

\begin{table}
	\footnotesize
	\setstretch{1.523} 
	\setlength\tabcolsep{5pt}
	\centering
	\caption{Auxiliary functions $A,B,C$ and $D$. An overall coefficient $m_{\Lambda_b}^2$ is implicit for the entries.}
	\begin{tabular*}{158mm}{c@{\extracolsep{\fill}}cccc}
		\toprule[1pt]
		\toprule[0.7pt]
		&A&B&C&D\\
		\toprule[0.7pt]
		$D_1$&$1-x_1^\prime$&$1+(x_1+x_3)(x_3^\prime-1)$&$x_2x_2^\prime$&$x_3x_3^\prime$\\
		$D_2$&$1-x_1^\prime$&$1+(x_1+x_2)(x_2^\prime-1)$&$x_2x_2^\prime$&$x_3x_3^\prime$\\
		$D_3$&$1-x_1^\prime$&$x_3(x_2^\prime+x_3^\prime)$&$(x_2+x_3)(x_2^\prime+x_3^\prime)$&$x_3x_3^\prime$\\
		$D_4$&$1-x_1^\prime$&$x_3^\prime(x_2+x_3)$&$(x_2+x_3)(x_2^\prime+x_3^\prime)$&$x_3x_3^\prime$\\
		$D_5$&$1-x_1^\prime$&$x_2^\prime(x_2+x_3)$&$x_2x_2^\prime$&$(x_2+x_3)(x_2^\prime+x_3^\prime)$\\
		$D_6$&$1-x_1^\prime$&$x_2(x_2^\prime+x_3^\prime)$&$x_2x_2^\prime$&$(x_2+x_3)(x_2^\prime+x_3^\prime)$\\
		$D_7$&$x_3(x_1^\prime+x_3^\prime)$&$(1-x_1)$&$x_2x_2^\prime$&$x_3x_3^\prime$\\
		$D_8$&$x_2(x_1^\prime+x_2^\prime)$&$(1-x_1)$&$x_2x_2^\prime$&$x_3x_3^\prime$\\
		$D_9$&$1-x_1$&$x_3(x_2^\prime+x_3^\prime)$&$(x_2+x_3)(x_2^\prime+x_3^\prime)$&$x_3x_3^\prime$\\
		$D_{10}$&$1-x_1$&$x_3^\prime(x_2+x_3)$&$(x_2+x_3)(x_2^\prime+x_3^\prime)$&$x_3x_3^\prime$\\
		$D_{11}$&$1-x_1$&$x_2(x_2^\prime+x_3^\prime)$&$x_2x_2^\prime$&$(x_2+x_3)(x_2^\prime+x_3^\prime)$\\
		$D_{12}$&$1-x_1$&$x_2^\prime(x_2+x_3)$&$x_2x_2^\prime$&$(x_2+x_3)(x_2^\prime+x_3^\prime)$\\
		$D_{13}$&$x_2(x_1^\prime+x_2^\prime)$&$1+(x_1+x_3)(x_3^\prime-1)$&$x_2x_2^\prime$&$x_3x_3^\prime$\\
		$D_{14}$&$x_3(x_1^\prime+x_3^\prime)$&$1+(x_1+x_2)(x_2^\prime-1)$&$x_2x_2^\prime$&$x_3x_3^\prime$\\
		\toprule[0.7pt]
		\toprule[1pt]
	\end{tabular*}\label{table:ABCD}
\end{table}

\begin{table}
	\tiny
	\setstretch{1.523} 
	\setlength\tabcolsep{-5pt}
	\centering
	\caption{The same as Table~\ref{table:hardkernelfunction} but for the diagrams $D_1, D_2, D_3, D_4$ and $D_5$ in Fig.~\ref{fig:feynman}.}
	\begin{tabular}{ccc}
		\toprule[1pt]
		\toprule[0.7pt]
		&twist-3&twist-4\\
		\toprule[0.7pt]
		$D_1$&&\\
		twist-2&$\psi_24(1-x_2)(V_1-A_1+2T_1)$&$r\psi_24(1-x_2)(P_1-A_3-S_1-T_3-T_7-V_3)$\\
		twist-$3^{+-}$&$\psi_3^{+-}2(x_3-x_1x_3)(V_1+A_1)$&$r\psi_3^{+-}2x_3(V_3-A_3)$\\
		twist-$3^{-+}$&$0$&$r\psi_3^{-+}2x_3(2T_2+T_3-T_7+S_1+P_1)$\\
		twist-4&$\psi_48(-x_2-x_3^\prime)T_1$&$r\psi_44(x_1-1)(1-x_2^\prime)(V_2-V_3-A_2-A_3)$\\
		\cline{2-3}
		&twist-5&twist-6\\
		\cline{2-3}
		twist-2&$r^2\psi_24(x_2+x_3^\prime)(-A_4-V_4)$&$r^3\psi_28(x_1^\prime x_2+x_3^\prime)T_6$\\
		twist-$3^{+-}$&$r^2\psi_3^{+-}2(x_1-1)(1-x_2^\prime)(T_4+2T_5-T_8+S_2+P_2)$&$0$\\
		twist-$3^{-+}$&$r^2\psi_3^{-+}2(x_1-1)(1-x_2^\prime)(V_4-A_4-T_8)$&$r^3\psi_3^{-+}2(1-x_2^\prime)(-V_6-A_6)$\\
		twist-4&$r^2\psi_44(1-x_2^\prime)(V_4-V_5+A_4+A_5+T_4+T_8+S_2-P_2)$&$0$\\
		\toprule[0.7pt]
		$D_2$&&\\
		twist2&$\psi_24(1-x_3)(V_1-A_1+2T_1)$&$r\psi_24(1-x_3)(-V_3-A_3-T_3-T_7-S_1+P_1)$\\
		twist$3^{+-}$&$0$&$r\psi_3^{+-}2(2T_2-V_2-A_2+(x_3+x_2^\prime)(T_3-T_7-S_1-P_1))$\\
		twist$3^{-+}$&$0$&$r\psi_3^{-+}2((-x_3-x_2^\prime)(V_2+A_2)+V_3-A_3+T_3-T_7-S_1-P_1)$\\
		twist4&$\psi_44(-x_3-x_2^\prime)(V_1-A_1)$&$r\psi_44(x_1^\prime x_3+x_2^\prime)(T_3+T_7+S_1-P_1)$\\
		\cline{2-3}
		&twist-5&twist-6\\
		\cline{2-3}
		twist-2&$r^2\psi_24(-x_3-x_2^\prime)(T_4+T_8+S_2-P_2)$&$r^3\psi_24(x_3x_1^\prime+x_2^\prime)(V_6-A_6)$\\
		twist-$3^{+-}$&$r^2\psi_3^{+-}2(x_1^\prime(T_8+S_2+P_2-V_4+A_4-T_4)+(x_1^\prime x_3+x_2^\prime)(V_5+A_5))$&$0$\\
		twist-$3^{-+}$&$r^2\psi_3^{-+}2(V_5+A_5-2T_5+x_1^\prime T_8+(-x_1^\prime x_3-x_2^\prime)(T_4-S_2-P_2))$&$0$\\
		twist-4&$r^2\psi_44(1-x_2^\prime)(V_4+A_4+T_4+T_8+S_2-P_2)$&$r^3\psi_44x_1^\prime(x_2^\prime-1)(V_6-A_6+2T_6)$\\
		\toprule[0.7pt]
		$D_3$&&\\
		twist2&$0$&$0$\\
		twist$3^{+-}$&$0$&$r\psi_3^{+-}2(1-x_1^\prime)(-V_2+V_3-A_2-A_3+2T_2+T_3-T_7-S_1-P_1)$\\
		twist$3^{-+}$&$0$&$r\psi_3^{-+}2x_3(V_2-V_3+A_2+A_3-2T_2-T_3+T_7+S_1+P_1)$\\
		twist4&$\psi_44x_3(V_1-A_1+2T_1)$&$r\psi_44x_1^\prime x_3(-V_3-A_3-T_3-T_7-S_1+P_1)$\\
		\cline{2-3}
		&twist-5&twist-6\\
		\cline{2-3}
		twist-2&$r^2\psi_24(x_1^\prime-1)(V_4+A_4+T_4+T_7+S_1-P_1)$&$r^3\psi_24(1-x_1^\prime)(V_6-A_6+2T_6)$\\
		twist-$3^{+-}$&$r^2\psi_3^{+-}2(1-x_1^\prime)(-V_4+V_5+A_4+A_5-T_4-2T_5+T_8+S_2+P_2)$&$0$\\
		twist-$3^{-+}$&$r^2\psi_3^{-+}2(x_1^\prime x_3(V_4-V_5-A_4-A_5+T_4+2T_5-S_2-P_2)+2(1-x_1^\prime))$&$0$\\
		twist-4&$0$&$0$\\
		\toprule[0.7pt]
		$D_4$&&\\
		twist2&$0$&$0$\\
		twist$3^{+-}$&$0$&$r\psi_3^{+-}2(x_3^\prime(V_2+A_2-2T_2)+(1-x_1)(V_3-A_3+T_3-T_7-S_1-P_1))$\\
		twist$3^{-+}$&$0$&$r\psi_3^{-+}2(x_3^\prime(V_2+A_2-2T_2)+(1-x_1)(V_3-A_3+T_3-T_7-S_1-P_1))$\\
		twist4&$\psi_44x_3^\prime(V_1-A_1+2T_1)$&$r\psi_44x_3^\prime(-V_3-A_3-T_3-T_7-S_1+P_1)$\\
		\cline{2-3}
		&twist-5&twist-6\\
		\cline{2-3}
		twist-2&$r^2\psi_24(1-x_1)(-V_4-A_4-T_4-T_8-S_2+P_2)$&$r^3\psi_24x_1^\prime(1-x_1)(V_6-A_6+2T_6)$\\
		twist-$3^{+-}$&$r^2\psi_3^{+-}2(x_3^\prime(V_4-A_4+T_4-T_8-S_2-P_2)+x_1^\prime(1-x_1)(V_5+A_5-2T_5))$&$0$\\
		twist-$3^{-+}$&$r^2\psi_3^{-+}2(x_3^\prime(V_4-A_4+T_4-T_8-S_2-P_2)+x_1^\prime(1-x_1)(V_5+A_5-2T_5))$&$0$\\
		twist-4&$0$&$0$\\
		\toprule[0.7pt]
		$D_5$&&\\
		twist2&$0$&$0$\\
		twist$3^{+-}$&$0$&$r\psi_3^{+-}2((1-x_1)(-V_2-A_2+2T_2)-x_2^\prime(-V_3+A_3-T_3+T_7+S_1+P_1))$\\
		twist$3^{-+}$&$0$&$r\psi_3^{-+}2((1-x_1)(-V_2-A_2+2T_2)-x_2^\prime(V_3-A_3+T_3-T_7-S_1-P_1))$\\
		twist4&$\psi_44x_2^\prime(V_1-A_1+2T_1)$&$r\psi_44x_2^\prime(-V_3-A_3-T_3-T_7-S_1+P_1)$\\
		\cline{2-3}
		&twist-5&twist-6\\
		\cline{2-3}
		twist-2&$r^2\psi_24(1-x_1)(-V_4-A_4-T_4-T_8-S_2+P_2)$&$r^3\psi_24x_1^\prime (1-x_1)(V_6-A_6+2T_6)$\\
		twist-$3^{+-}$&$r^2\psi_3^{+-}2(x_1^\prime(1-x_1)(-V_4+A_4-T_4+T_8+S_2+P_2)-x_2^\prime(V_5+A_5-2T_5))$&$0$\\
		twist-$3^{-+}$&$r^2\psi_3^{-+}2(x_1^\prime(1-x_1)(-V_4+A_4-T_4+T_8+S_2+P_2)-2x_2^\prime(V_5+A_5-2T_5))$&$0$\\
		twist-4&$0$&$0$\\
		\toprule[0.7pt]
		\toprule[1pt]
	\end{tabular}\label{table:hardfunctions1}
\end{table}

\begin{table}
	\tiny
	\setstretch{1.523} 
	\setlength\tabcolsep{0pt}
	\centering
	\caption{The same as Table~\ref{table:hardkernelfunction} but for the diagrams $D_6, D_7, D_8, D_9$ and $D_{10}$ in Fig.~\ref{fig:feynman}.}
	\begin{tabular}{ccc}
		\toprule[1pt]
		\toprule[0.7pt]
		&twist3&twist4\\
		\toprule[0.7pt]
		$D_6$&&\\
		twist2&$0$&$0$\\
		twist$3^{+-}$&$0$&$r\psi_3^{+-}2x_2(V_2-V_3+A_2+A_3-2T_2-T_3+T_7+S_1+P_1)$\\
		twist$3^{-+}$&$0$&$r\psi_3^{-+}2(1-x_1^\prime)(-V_2+V_3-A_2-A_3+2T_2+T_3-T_7-S_1-P_1)$\\
		twist4&$\psi_44x_2(V_1-A_1+2T_1)$&$r\psi_44x_1^\prime x_2(-V_3-A_3-T_3-T_7-S_1+P_1)$\\
		\cline{2-3}
		&twist-5&twist-6\\
		\cline{2-3}
		twist-2&$r^2\psi_24(1-x_1^\prime)(-V_4-A_4-T_4-T_8-S_2+P_2)$&$r^3\psi_24(1-x_1^\prime)(V_6-A_6+2T_6)$\\
		twist-$3^{+-}$&$r^2\psi_3^{+-}2x_1^\prime x_2(V_4-V_5-A_4-A_5+T_4+2T_5-T_8-S_2-P_2)$&$0$\\
		twist-$3^{-+}$&$r^2\psi_3^{-+}2(1-x_1^\prime)(-V_4+V_5+A_4+A_5-T_4-2T_5-T_8+S_2+P_2)$&$0$\\
		twist-4&$0$&$0$\\
		\toprule[0.7pt]
		$D_7$&&\\
		twist2&$0$&$r\psi_24(x_1-1)x_3(-V_2+V_3+A_2+A_3+T_3+T_7+S_1-P_1)$\\
		twist$3^{+-}$&$\psi_3^{+-}2(x_3-x_1x_3)(V_1+A_1)$&$r\psi_3^{+-}2x_3(V_3-A_3)$\\
		twist$3^{-+}$&$0$&$r\psi_3^{-+}2x_3(2T_2+T_3-T_7+S_1+P_1)$\\
		twist4&$\psi_48x_3(-T_1)$&$r\psi_44(x_1-1)(1-x_2^\prime)(V_2-V_3-A_2-A_3)$\\
		\cline{2-3}
		&twist-5&twist-6\\
		\cline{2-3}
		twist-2&$r^2\psi_24x_3(-V_4+V_5-A_4-A-5)$&$r^3\psi_28(1-x_1)(1-x_2^\prime)T_6$\\
		twist-$3^{+-}$&$r^2\psi_3^{+-}2(x_1-1)(1-x_2^\prime)(T_4+2T_5-T_8+S_2+P_2)$&$0$\\
		twist-$3^{-+}$&$r^2\psi_3^{-+}2(x_1-1)(1-x_2^\prime)(V_4-A_4-T_8)$&$r^3\psi_3^{-+}2(1-x_2^\prime)(-V_6-A_6)$\\
		twist-4&$r^2\psi_44(1-x_2^\prime)(V_4-V_5+A_4+A_5+T_4+T_8+S_2-P_2)$&$0$\\
		\toprule[0.7pt]
		$D_8$&&\\
		twist2&$0$&$r\psi_24(x_1-1)x_2(V_3+A_3+2S_1-2P_1)$\\
		twist$3^{+-}$&$0$&$r\psi_3^{+-}4x_2(-S_1-P_1)$\\
		twist$3^{-+}$&$\psi_3^{-+}2(x_1-1)x_2(V_1+A_1)$&$r\psi_3^{-+}2x_2(-V_2-A_2)$\\
		twist4&$\psi_44x_2(-V_1+A_1)$&$r\psi_48(x_1-1)(1-x_3^\prime)(-S_1+P_1)$\\
		\cline{2-3}
		&twist5&twist6\\
		\cline{2-3}
		twist2&$r^2\psi_28x_2(-S_2+P_2)$&$r^3\psi_24(x_1-1)(1-x_2^\prime)(-V_6+A_6)$\\
		twist$3^{+-}$&$r^2\psi_3^{+-}2(x_1-1)(1-x_3^\prime)(-V_5-A_5)$&$r^3\psi_3^{+-}2(1-x_3^\prime)(V_6+A_6)$\\
		twist$3^{-+}$&$r^2\psi_3^{-+}4(x_1-1)(1-x_3^\prime)(-S_2-P_2)$&$0$\\
		twist4&$r^2\psi_44(1-x_3^\prime)(V_4+A_4+2S_2-2P_2)$&$0$\\
		\toprule[0.7pt]
		$D_9$&&\\
		twist2&$0$&$r\psi_24(x_1-1)x_3(-T_3-T_7+S_1-P_1)$\\
		twist$3^{+-}$&$0$&$0$\\
		twist$3^{-+}$&$0$&$r\psi_3^{-+}2x_3(V_2-V_3+A2+A_3-2T_2-T_3+T_7+S_1+P_2)$\\
		twist4&$\psi_44x_3(V_1-A_1+2T_1)$&$r\psi_44(x_1-1)(1-x_1^\prime)(V_2-A_2)$\\
		\cline{2-3}
		&twist5&twist6\\
		\cline{2-3}
		twist2&$r^2\psi_24x_3(V_5-A_5)$&$r^3\psi_24(x_1-1)(1-x_1^\prime)(V_6-A_6+2T_6)$\\
		twist$3^{+-}$&$r^2\psi_3^{+-}2(x_1-1)(1-x_1^\prime)(-V_4+V_5+A_4+A_5-T_4-2T_5+T_8+S_2+P_2)$&$0$\\
		twist$3^{-+}$&$r^2\psi_3^{-+}2(x_1-1)(1-x_1^\prime)T_8$&$0$\\
		twist4&$r^2\psi_44(1-x_1^\prime)(-T_4-T_8+S_2-P_2)$&$0$\\
		\toprule[0.7pt]
		$D_{10}$&&\\
		twist2&$0$&$r\psi_24(x_1-1)(1-x_1)(-V_2+A_2)$\\
		twist$3^{+-}$&$\psi_3^{+-}2(x_1-1)(1-x_1)(-V_1-A_1)$&$r\psi_3^{+-}2(1-x_1)(V_3-A_3-2S_1-2P_1)$\\
		twist$3^{-+}$&$\psi_3^{-+}2(x_1-1)(1-x_1)(-V_1-A_1)$&$r\psi_3^{-+}2(1-x_1)(V_3-A_3-2S_2-2P_1)$\\
		twist4&$0$&$r\psi_44(x_1-1)x_3^\prime(-V_3-A_3-2S_1+2P_1)$\\
		\cline{2-3}
		&twist5&twist6\\
		\cline{2-3}
		twist2&$r^2\psi_24(1-x_1)(-V_4-A_4-2S_2+2P_2)$&$0$\\
		twist$3^{+-}$&$r^2\psi_3^{+-}2(x_1-1)x_3^\prime(V_4+A_4-S_2-P_2)$&$r^3\psi_3^{+-}2x_3^\prime(-V_6-A_6)$\\
		twist$3^{-+}$&$r^2\psi_3^{-+}2(x_1-1)x_3^\prime(V_4+A_4-S_2-P_2)$&$r^3\psi_3^{-+}2x_3^\prime(-V_6-A-6)$\\
		twist4&$r^2\psi_44x_3^\prime(-V_5+A_5)$&$0$\\
		\toprule[0.7pt]
		\toprule[1pt]
	\end{tabular}\label{table:hardfunctions2}
\end{table}

\begin{table}
	\tiny
	\setstretch{1.523} 
	\setlength\tabcolsep{3pt}
	\centering
	\caption{The same as Table~\ref{table:hardkernelfunction} but for the diagrams $D_{11}, D_{12}, D_{13}$ and $D_{14}$ in Fig.~\ref{fig:feynman}.}
	\begin{tabular}{ccc}
		\toprule[1pt]
		\toprule[0.7pt]
		&twist3&twist4\\
		\toprule[0.7pt]
		$D_{11}$&&\\
		twist2&$0$&$r\psi_24(x_1-1)x_2(-V_2+A_2)$\\
		twist$3^{+-}$&$0$&$r\psi_3^{+-}2x_2(V_2-V_3+A_2+A_3-2T_2-T_3+T_7+S_1+P_1)$\\
		twist$3^{-+}$&$0$&$0$\\
		twist4&$\psi_44x_2(V_1-A_1+2T_1)$&$r\psi_44(x_1-1)(1-x_1^\prime)(T_3+T_7-S_1+P_1)$\\
		\cline{2-3}
		&twist5&twist6\\
		\cline{2-3}
		twist2&$r^2\psi_24x_2(T_4+T_8-S_2+P_2)$&$r^3\psi_24(x_1-1)(1-x_1^\prime)(V_6-A_6+2T_6)$\\
		twist$3^{+-}$&$0$&$0$\\
		twist$3^{-+}$&$r^2\psi_3^{-+}2(x_1-1)(1-x_1^\prime)(-V_4+V_5+A_4+A_5-T_4-2T_5+S_2+P_2)$&$0$\\
		twist4&$r^2\psi_44(1-x_1^\prime)(-V_5+A_5)$&$0$\\
		\toprule[0.7pt]
		$D_{12}$&&\\
		twist2&$0$&$r\psi_24(x_1-1)(1-x_1)(-T_3-T_7+S_1-P_1)$\\
		twist$3^{+-}$&$\psi_3^{+-}2(x_1-1)(1-x_1)(V_1+A_1)$&$r\psi_3^{+-}2(1-x_1)(-V_2-A_2+2T_2+T_3-T_7+S_1+P_1)$\\
		twist$3^{-+}$&$\psi_3^{-+}2(x_1-1)(1-x_1)(V_1+A_1)$&$r\psi_3^{-+}2(1-x_1)(-V_2-A_2+2T_2+T_3-T_7+S_1+P_1)$\\
		twist4&$0$&$r\psi_44(x_1-1)x_2^\prime(V_2-V_3-A_2-A_3-T_3-T_7-S_1+P_1)$\\
		\cline{2-3}
		&twist5&twist6\\
		\cline{2-3}
		twist2&$r^2\psi_24(1-x_1)(-V_4+V_5-A_4-A_5-T_4-T_8-S_2+P_2)$&$0$\\
		twist$3^{+-}$&$r^2\psi_3^{+-}2(x_2^\prime-x_1x_2^\prime)(V_5+A_5+T_4+2T_5+T_8+S_2+P_2)$&$r^3\psi_3^{+-}2x_2^\prime(V_6+A_6)$\\
		twist$3^{-+}$&$r^2\psi_3^{-+}2(x_2^\prime-x_1x_2^\prime)(V_5+A_5+T_4+2T_5+T_8+S_2+P_2)$&$r^3\psi_3^{-+}2x_2^\prime(V_6+A_6)$\\
		twist4&$r^2\psi_43x_2^\prime(-T_4-T_8+S_2-P_2)$&$0$\\
		\toprule[0.7pt]
		$D_{13}$&&\\
		twist2&$0$&$r\psi_24x_2(V_3+A_3+2S_1-2P_1)$\\
		twist$3^{+-}$&$0$&$0$\\
		twist$3^{-+}$&$\psi_3^{-+}2x_2(V_1+A_1)$&$r\psi_3^{-+}2(1-x_2)(1-x_3^\prime)(V_3-A_3-2S_1-2P_1)$\\
		twist4&$0$&$r\psi_44x_2(x_3^\prime-1)(V_2-A_2)$\\
		\cline{2-3}
		&twist5&twist6\\
		\cline{2-3}
		twist2&$r^2\psi_24(1-x_2)(1-x_3^\prime)(-V_5+A_5)$&$0$\\
		twist$3^{+-}$&$r^2\psi_3^{+-}2(x_2-x_2x_3^\prime)(V_4+A_4+S_2+P_2)$&$r^3\psi_3^{+-}2(1-x_3^\prime)(V_6+A_6)$\\
		twist$3^{-+}$&$0$&$0$\\
		twist4&$r^2\psi_44(1-x_3^\prime)(V_4+A_4+2S_2-2P_2)$&$0$\\
		\toprule[0.7pt]
		$D_{14}$&&\\
		twist2&$0$&$r\psi_24x_3(-V_2+V_3+A_2+A_3+T_3+T_7+S_1-P_1)$\\
		twist$3^{+-}$&$\psi_3^{+-}2x_3(-V_1-A_1)$&$r\psi_3^{+-}2(1-x_3)(1-x_2^\prime)(-V_2-A_2+2T_2+T_3-T_7+S_1+P_1)$\\
		twist$3^{-+}$&$0$&$0$\\
		twist4&$0$&$r\psi_44(x_2^\prime-1)x_3(T_3+T_7-S_1+P_1)$\\
		\cline{2-3}
		&twist5&twist6\\
		\cline{2-3}
		twist2&$r^2\psi_24(1-x_3)(1-x_2^\prime)(-T_4-T_8+S_2-P_2)$&$0$\\
		twist$3^{+-}$&$0$&$0$\\
		twist$3^{-+}$&$r^2\psi_3^{-+}2(x_1^\prime-1)(V_5+A_5+T_4-2T_5+S_2+P_2)$&$r^3\psi_3^{-+}2(1-x_2^\prime)(-V_6-A_6)$\\
		twist4&$r^2\psi_44(1-x_2^\prime)(V_4-V_5+A_4+A_5+T_4+T_7+S_1-P_2)$&$0$\\
		\toprule[0.7pt]
		\toprule[1pt]
	\end{tabular}\label{table:hardfunctions3}
\end{table}


\begin{thebibliography}{100}
	
	
	\bibitem{LHCb:2016yco}
	R.~Aaij \textit{et al.} [LHCb],
	``Measurement of matter-antimatter differences in beauty baryon decays,''
	Nature Phys. \textbf{13}, 391 (2017)
	[arXiv:1609.05216 [hep-ex]].
	
	\bibitem{LHCb:2018fly}
	R.~Aaij \textit{et al.} [LHCb Collab.],
	``Search for $C\!P$ violation in $\Lambda^0_b \to p K^-$ and $\Lambda^0_b \to p \pi^-$ decays,''
	Phys. Lett. B \textbf{787}, 124 (2018)
	[arXiv:1807.06544 [hep-ex]].
	
	\bibitem{LHCb:2018fpt}
	R.~Aaij \textit{et al.} [LHCb],
	``Search for CP violation using triple product asymmetries in $\Lambda^{0}_{b}\to p K^{-}\pi^{+}\pi^{-}$, $\Lambda^{0}_{b}\to pK^{-}K^{+}K^{-}$ and $\Xi^{0}_{b}\to pK^{-}K^{-}\pi^{+}$ decays,''
	JHEP \textbf{08}, 039 (2018)
	[arXiv:1805.03941 [hep-ex]].
	
	\bibitem{LHCb:2019jyj}
	R.~Aaij \textit{et al.} [LHCb],
	``Measurements of $CP$ asymmetries in charmless four-body $\Lambda_b^0$ and $\Xi_b^0$ decays,''
	Eur. Phys. J. C \textbf{79}, 745 (2019)
	[arXiv:1903.06792 [hep-ex]].
	
	\bibitem{Mannel:1990vg}
	T.~Mannel, W.~Roberts and Z.~Ryzak,``Baryons in the heavy quark effective theory,''
	Nucl. Phys. B \textbf{355}, 38 (1991).
	
	\bibitem{Hussain:1992rb}
	F.~Hussain, D.~S.~Liu, M.~Kramer, J.~G.~Korner and S.~Tawfiq,
	``General analysis of weak decay form-factors in heavy to heavy and heavy to light baryon transitions,''
	Nucl. Phys. B \textbf{370}, 259 (1992).
	
	\bibitem{Isgur:1990pm}
	N.~Isgur and M.~B.~Wise, ``Heavy baryon weak form-factors,''
	Nucl. Phys. B \textbf{348}, 276 (1991).
	
	\bibitem{Georgi:1990cx}
	H.~Georgi, ``Comment on heavy baryon weak form-factors,''
	Nucl. Phys. B \textbf{348}, 293 (1991).
	
	\bibitem{Bauer:2000yr}
	C.~W.~Bauer, S.~Fleming, D.~Pirjol and I.~W.~Stewart,
	``An Effective field theory for collinear and soft gluons: Heavy to light decays,''
	Phys. Rev. D \textbf{63}, 114020 (2001)
	[arXiv:hep-ph/0011336 [hep-ph]].
	
	\bibitem{Bauer:2001yt}
	C.~W.~Bauer, D.~Pirjol and I.~W.~Stewart,
	``Soft collinear factorization in effective field theory,''
	Phys. Rev. D \textbf{65}, 054022 (2002)
	[arXiv:hep-ph/0109045 [hep-ph]].
	
	\bibitem{Beneke:1999br}
	M.~Beneke, G.~Buchalla, M.~Neubert and C.~T.~Sachrajda,
	``QCD factorization for $B \to \pi \pi$ decays: strong phases and CP violation in the heavy quark limit,''
	Phys. Rev. Lett. \textbf{83}, 1914 (1999)
	[arXiv:hep-ph/9905312 [hep-ph]].
	
	\bibitem{Beneke:2000ry}
	M.~Beneke, G.~Buchalla, M.~Neubert and C.~T.~Sachrajda,
	``QCD factorization for exclusive, nonleptonic B meson decays: General arguments and the case of heavy light final states,'' Nucl. Phys. B \textbf{591}, 313 (2000)
	[arXiv:hep-ph/0006124 [hep-ph]].
	
	\bibitem{Beneke:2001ev}
	M.~Beneke, G.~Buchalla, M.~Neubert and C.~T.~Sachrajda,
	``QCD factorization in $B \to \pi K, \pi \pi$ decays and extraction of Wolfenstein parameters,''
	Nucl. Phys. B \textbf{606}, 245 (2001)
	[arXiv:hep-ph/0104110 [hep-ph]].
	
	\bibitem{Beneke:2003zv}
	M.~Beneke and M.~Neubert, ``QCD factorization for $B \to PP$ and $B \to PV$ decays,''
	Nucl. Phys. B \textbf{675}, 333 (2003)
	[arXiv:hep-ph/0308039 [hep-ph]].
	
	\bibitem{Keum:2000wi}
	Y.~Y.~Keum, H.~n.~Li and A.~I.~Sanda,
	``Penguin enhancement and $B \to K \pi$ decays in perturbative QCD,''
	Phys. Rev. D \textbf{63}, 054008 (2001)
	[arXiv:hep-ph/0004173 [hep-ph]].
	
	\bibitem{Lu:2000em}
	C.~D.~L\"u, K.~Ukai and M.~Z.~Yang,
	``Branching ratio and CP violation of $B\to\pi \pi$ decays in perturbative QCD approach,''
	Phys. Rev. D \textbf{63}, 074009 (2001)
	[arXiv:hep-ph/0004213 [hep-ph]].
	
	\bibitem{Keum:2000ph}
	Y.~Y.~Keum, H.~n.~Li and A.~I.~Sanda, ``Fat penguins and imaginary penguins in perturbative QCD,''
	Phys. Lett. B \textbf{504}, 6 (2001)
	[arXiv:hep-ph/0004004 [hep-ph]].
	
	\bibitem{Chau:1987tk}
	L.~L.~Chau and H.~Y.~Cheng,
	``Analysis of exclusive two-body decays of charm mesons using the quark diagram scheme,''
	Phys. Rev. D \textbf{36}, 137 (1987).
	
	\bibitem{Cheng:2010ry}
	H.~Y.~Cheng and C.~W.~Chiang, ``Two-body hadronic charmed meson decays,''
	Phys. Rev. D \textbf{81}, 074021 (2010)
	[arXiv:1001.0987 [hep-ph]].
	
	\bibitem{Li:2012cfa}
	H.~n.~Li, C.~D.~L\"u and F.~S.~Yu,
	``Branching ratios and direct CP asymmetries in $D\to PP$ decays,''
	Phys. Rev. D \textbf{86}, 036012 (2012)
	[arXiv:1203.3120 [hep-ph]].
	
	%
	\bibitem{Cheng:2004ru}
	H.~Y.~Cheng, C.~K.~Chua and A.~Soni, ``Final state interactions in hadronic B decays,''
	Phys. Rev. D \textbf{71}, 014030 (2005)
	[arXiv:hep-ph/0409317 [hep-ph]].
	
	\bibitem{Yu:2017zst}
	F.~S.~Yu, H.~Y.~Jiang, R.~H.~Li, C.~D.~L\"u, W.~Wang and Z.~X.~Zhao,
	``Discovery potentials of doubly charmed baryons,''
	Chin. Phys. C \textbf{42}, 051001 (2018)
	[arXiv:1703.09086 [hep-ph]].
	
	
	\bibitem{Han:2021azw}
	J.~J.~Han, H.~Y.~Jiang, W.~Liu, Z.~J.~Xiao and F.~S.~Yu,
	``Rescattering mechanism of weak decays of double-charm baryons,''
	Chin. Phys. C \textbf{45}, 053105 (2021)
	[arXiv:2101.12019 [hep-ph]].
	
	\bibitem{Savage:1989ub}
	M.~J.~Savage and M.~B.~Wise, ``SU(3) predictions for nonleptonic B meson decays,''
	Phys. Rev. D \textbf{39}, 3346 (1989) [erratum: Phys. Rev. D \textbf{40}, 3127 (1989)].
	
	\bibitem{He:2018php}
	X.~G.~He and W.~Wang,
	``Flavor SU(3) topological diagram and irreducible representation amplitudes for heavy meson charmless hadronic decays: mismatch and equivalence,''
	Chin. Phys. C \textbf{42}, 103108 (2018)
	[arXiv:1803.04227 [hep-ph]].
	
	\bibitem{Wang:2020gmn}
	D.~Wang, C.~P.~Jia and F.~S.~Yu,
	``A self-consistent framework of topological amplitude and its $SU(N)$ decomposition,''
	JHEP \textbf{09}, 126 (2020)
	[arXiv:2001.09460 [hep-ph]].
	
	\bibitem{Hsiao:2014mua}
	Y.~K.~Hsiao and C.~Q.~Geng, ``Direct CP violation in $\Lambda_b$ decays,''
	Phys. Rev. D \textbf{91}, 116007 (2015)
	[arXiv:1412.1899 [hep-ph]].
	
	\bibitem{Hsiao:2017tif}
	Y.~K.~Hsiao, Y.~Yao and C.~Q.~Geng, ``Charmless two-body anti-triplet $b$-baryon decays,''
	Phys. Rev. D \textbf{95}, 093001 (2017)
	[arXiv:1702.05263 [hep-ph]].
	
	\bibitem{Geng:2021nkl}
	C.~Q.~Geng, C.~W.~Liu and T.~H.~Tsai,
	``Non-leptonic two-body decays of $\Lambda^0_b$ in light-front quark model,''
	Phys. Lett. B \textbf{815}, 136125 (2021)
	[arXiv:2102.01552 [hep-ph]].
	
	\bibitem{Zhu:2016bra}
	J.~Zhu, H.~W.~Ke and Z.~T.~Wei,
	``The decay of $\Lambda _b\to p~K^-$ in QCD factorization approach,''
	Eur. Phys. J. C \textbf{76}, 284 (2016)
	[arXiv:1603.02800 [hep-ph]].
	
	\bibitem{Lu:2009cm}
	C.~D.~Lu, Y.~M.~Wang, H.~Zou, A.~Ali and G.~Kramer,
	``Anatomy of the pQCD approach to the baryonic decays $\Lambda_b \to p \pi, p K$,''
	Phys. Rev. D \textbf{80}, 034011 (2009)
	[arXiv:0906.1479 [hep-ph]].
	
	\bibitem{Cheng:2014rfa}
	H.~Y.~Cheng, C.~W.~Chiang and A.~L.~Kuo,
	``Updating $B \to PP,VP$ decays in the framework of flavor symmetry,''
	Phys. Rev. D \textbf{91}, 014011 (2015)
	[arXiv:1409.5026 [hep-ph]].
	
	\bibitem{Qin:2021tve}
	Q.~Qin, C.~Wang, D.~Wang and S.~H.~Zhou,
	``The factorization-assisted topological-amplitude approach and its applications,''
	[arXiv:2111.14472 [hep-ph]].
	
	\bibitem{Li:2007bpa}
	H.~n.~Li, ``Factorization approaches to B meson decays,''
	eConf \textbf{C070512}, 011 (2007)
	[arXiv:0707.1294 [hep-ph]].
	
	\bibitem{Li:2003yj}
	H.~n.~Li, ``QCD aspects of exclusive B meson decays,''
	Prog. Part. Nucl. Phys. \textbf{51}, 85 (2003)
	[arXiv:hep-ph/0303116 [hep-ph]].
	
	
	\bibitem{Belle:2004mad}
	K.~Abe \textit{et al.} [Belle],
	``Observation of large $CP$ violation and evidence for direct $CP$ violation in $B^0 \to \pi^+\pi^-$ decays,'' Phys. Rev. Lett. \textbf{93}, 021601 (2004)
	[arXiv:hep-ex/0401029 [hep-ex]].
	
	\bibitem{BaBar:2004gyj}
	B.~Aubert \textit{et al.} [BaBar],
	``Observation of direct CP violation in $B^0 \to K^+ \pi^-$ decays,''
	Phys. Rev. Lett. \textbf{93}, 131801 (2004)
	[arXiv:hep-ex/0407057 [hep-ex]].
	
	\bibitem{Belle:2004nch}
	Y.~Chao \textit{et al.} [Belle],
	``Evidence for direct CP violation in $B^0 \to K^+ \pi^-$ decays,''
	Phys. Rev. Lett. \textbf{93}, 191802 (2004)
	[arXiv:hep-ex/0408100 [hep-ex]].
	
	\bibitem{Leibovich:2003tw}
	A.~K.~Leibovich, Z.~Ligeti, I.~W.~Stewart and M.~B.~Wise,
	``Predictions for nonleptonic $\Lambda_b$ and $\Theta_b$ decays,''
	Phys. Lett. B \textbf{586}, 337 (2004)
	[arXiv:hep-ph/0312319 [hep-ph]].
	
	
	\bibitem{Li:1992ce}
	H.~n.~Li, ``Sudakov suppression and the proton form-factor in QCD,''
	Phys. Rev. D \textbf{48}, 4243 (1993)
	
	\bibitem{Shih:1998pb}
	H.~H.~Shih, S.~C.~Lee and H.~n.~Li,
	``The $\Lambda_b\to p$ lepton anti-neutrino decay in perturbative QCD,''
	Phys. Rev. D \textbf{59}, 094014 (1999)
	[arXiv:hep-ph/9810515 [hep-ph]].
	
	\bibitem{Shih:1999yh}
	H.~H.~Shih, S.~C.~Lee and H.~n.~Li,
	``Applicability of perturbative QCD to $\Lambda_b \to \Lambda_c$ decays,''
	Phys. Rev. D \textbf{61}, 114002 (2000)
	[arXiv:hep-ph/9906370 [hep-ph]].
	
	\bibitem{Guo:2005qa}
	P.~Guo, H.~W.~Ke, X.~Q.~Li, C.~D.~Lu and Y.~M.~Wang,
	``Diquarks and the semi-leptonic decay of $\Lambda_b$ in the hybrid scheme,''
	Phys. Rev. D \textbf{75}, 054017 (2007)
	[arXiv:hep-ph/0501058 [hep-ph]].
	
	\bibitem{He:2006ud}
	X.~G.~He, T.~Li, X.~Q.~Li and Y.~M.~Wang,
	``PQCD calculation for $\Lambda_b \to \Lambda \gamma$ in the standard model,''
	Phys. Rev. D \textbf{74}, 034026 (2006)
	[arXiv:hep-ph/0606025 [hep-ph]].
	
	\bibitem{Chou:2001bn}
	C.~H.~Chou, H.~H.~Shih, S.~C.~Lee and H.~n.~Li,
	``$\Lambda_b \to \Lambda J/\psi$ decay in perturbative QCD,''
	Phys. Rev. D \textbf{65}, 074030 (2002)
	[arXiv:hep-ph/0112145 [hep-ph]].
	
	\bibitem{Zhang:2022iun}
	C.~Q.~Zhang, J.~M.~Li, M.~K.~Jia and Z.~Rui,
	Phys. Rev. D \textbf{105}, 073005 (2022)
	[arXiv:2202.09181 [hep-ph]].
	
	
	\bibitem{Detmold:2015aaa}
	W.~Detmold, C.~Lehner and S.~Meinel,
	``$\Lambda_b \to p \ell^- \bar{\nu}_\ell$ and $\Lambda_b \to \Lambda_c \ell^- \bar{\nu}_\ell$ form factors from lattice QCD with relativistic heavy quarks,''
	Phys. Rev. D \textbf{92}, 034503 (2015)
	[arXiv:1503.01421 [hep-lat]].
	
	\bibitem{Zhang:2021oja}
	Q.~A.~Zhang, J.~Hua, F.~Huang, R.~Li, Y.~Li, C.~D.~L\"u, P.~Sun, W.~Sun, W.~Wang and Y.~B.~Yang,
	``$\Xi_c\to \Xi$ Form Factors and $\Xi_c\to \Xi \ell^+\nu_{\ell}$ Decay Rates From Lattice QCD,''
	Chin. Phys. C \textbf{46}, 011002 (2022)
	[arXiv:2103.07064 [hep-lat]].
	
	
	\bibitem{Huang:1998rq}
	C.~S.~Huang, C.~F.~Qiao and H.~G.~Yan,
	``Decay $\Lambda_b\to p$ lepton anti-neutrino in QCD sum rules,''
	Phys. Lett. B \textbf{437}, 403 (1998)
	[arXiv:hep-ph/9805452 [hep-ph]].
	
	\bibitem{Wang:2009hra}
	Y.~M.~Wang, Y.~L.~Shen and C.~D.~Lu,
	``$\Lambda_b \to p, \Lambda$ transition form factors from QCD light-cone sum rules,''
	Phys. Rev. D \textbf{80}, 074012 (2009)
	[arXiv:0907.4008 [hep-ph]].
	
	\bibitem{Wang:2015ndk}
	Y.~M.~Wang and Y.~L.~Shen,
	``Perturbative Corrections to $\Lambda_b \to \Lambda$ form factors from QCD light-cone sum rules,''
	JHEP \textbf{02}, 179 (2016)
	[arXiv:1511.09036 [hep-ph]].
	
	\bibitem{Jaus:1999zv}
	W.~Jaus, ``Covariant analysis of the light front quark model,''
	Phys. Rev. D \textbf{60}, 054026 (1999).
	
	\bibitem{Wang:2017mqp}
	W.~Wang, F.~S.~Yu and Z.~X.~Zhao,
	``Weak decays of doubly heavy baryons: the $1/2\rightarrow 1/2$ case,''
	Eur. Phys. J. C \textbf{77}, 781 (2017)
	[arXiv:1707.02834 [hep-ph]].
	
	\bibitem{Li:2021qod}
	Y.~S.~Li, X.~Liu and F.~S.~Yu,
	``Revisiting semileptonic decays of $\Lambda_{b(c)}$ supported by baryon spectroscopy,''
	Phys. Rev. D \textbf{104}, 013005 (2021)
	[arXiv:2104.04962 [hep-ph]].
	
	
	\bibitem{Wang:2011uv}
	W.~Wang, ``Factorization of heavy-to-light baryonic transitions in SCET,''
	Phys. Lett. B \textbf{708}, 119 (2012) 
	[arXiv:1112.0237 [hep-ph]].
	
	\bibitem{Feldmann:2011xf}
	T.~Feldmann and M.~W.~Y.~Yip,
	``Form factors for $\Lambda_b \to \Lambda$ transitions in the  soft-collinear effective theory,''
	Phys. Rev. D \textbf{85}, 014035 (2012) [erratum: Phys. Rev. D \textbf{86}, 079901 (2012)]
	[arXiv:1111.1844 [hep-ph]].
	
	\bibitem{Li:2014xda}
	H.~n.~Li and Y.~M.~Wang, ``Non-dipolar Wilson links for transverse-momentum-dependent wave functions,''
	JHEP \textbf{06}, 013 (2015) 
	[arXiv:1410.7274 [hep-ph]].
	
	\bibitem{JHEP02-008}
	H.~n.~Li, Y.~L.~Shen and Y.~M.~Wang,
	``Resummation of rapidity logarithms in $B$ meson wave functions,''
	JHEP \textbf{02}, 008 (2013) [arXiv:1210.2978 [hep-ph]].
	
	\bibitem{Liu:2020upy}
	X.~Liu, H.~n.~Li and Z.~J.~Xiao,
	``Next-to-leading-logarithm $k_T$ resummation for $B_c\to J/\psi$ decays,''
	Phys. Lett. B \textbf{811}, 135892 (2020)
	[arXiv:2006.12786 [hep-ph]].
	
	
	\bibitem{Bolz:1994hb}
	J.~Bolz, R.~Jakob, P.~Kroll, M.~Bergmann and N.~G.~Stefanis,
	``A Critical analysis of the proton form-factor with Sudakov suppression and intrinsic transverse momentum,'' Z. Phys. C \textbf{66}, 267 (1995)
	[arXiv:hep-ph/9405340 [hep-ph]].
	
	\bibitem{Kundu:1998gv}
	B.~Kundu, H.~n.~Li, J.~Samuelsson and P.~Jain,
	``The Perturbative proton form-factor reexamined,''
	Eur. Phys. J. C \textbf{8}, 637 (1999)
	[arXiv:hep-ph/9806419 [hep-ph]].
	
	\bibitem{Li:1994cka}
	H.~n.~Li and H.~L.~Yu,``Extraction of $V_{ub}$ from decay $B \to \pi\ell\nu$,''
	Phys. Rev. Lett. \textbf{74}, 4388 (1995)
	[arXiv:hep-ph/9409313 [hep-ph]].
	
	\bibitem{Li:1994iu}
	H.~n.~Li and H.~L.~Yu, ``Perturbative QCD analysis of $B$ meson decays,''
	Phys. Rev. D \textbf{53}, 2480 (1996)
	[arXiv:hep-ph/9411308 [hep-ph]].
	
	\bibitem{Li:2001ay}
	H.~n.~Li, ``Threshold resummation for exclusive B meson decays,''
	Phys. Rev. D \textbf{66}, 094010 (2002)
	[arXiv:hep-ph/0102013 [hep-ph]].
	
	\bibitem{Zhang:2020qaz}
	Z.~Q.~Zhang and H.~n.~Li,
	``Next-to-leading-logarithm threshold resummation for exclusive B meson decays,''
	Eur. Phys. J. C \textbf{81}, 595 (2021)
	[arXiv:2007.11173 [hep-ph]].
	
	\bibitem{Ball:2008fw}
	P.~Ball, V.~M.~Braun and E.~Gardi,
	``Distribution amplitudes of the $\Lambda_b$ baryon in QCD,''
	Phys. Lett. B \textbf{665}, 197 (2008)
	[arXiv:0804.2424 [hep-ph]].
	
	
	
	\bibitem{Bell:2013tfa}
	G.~Bell, T.~Feldmann, Y.~M.~Wang and M.~W.~Y.~Yip,
	``Light-cone distribution amplitudes for heavy-quark hadrons,''
	JHEP \textbf{11}, 191 (2013)
	[arXiv:1308.6114 [hep-ph]].
	
	
	\bibitem{Ali:2012zza}
	A.~Ali, C.~Hambrock and A.~Y.~Parkhomenko, ``Light-cone wave functions of heavy baryons,''
	Theor. Math. Phys. \textbf{170}, 2 (2012).
	
	\bibitem{Groote:1997yr}
	S.~Groote, J.~G.~Korner and O.~I.~Yakovlev,
	``An analysis of diagonal and nondiagonal QCD sum rules for heavy baryons at next-to-leading order in $\alpha_s$,''
	Phys. Rev. D \textbf{56}, 3943 (1997)
	[arXiv:hep-ph/9705447 [hep-ph]].
	
	
	\bibitem{Groote:1996em}
	S.~Groote, J.~G.~Korner and O.~I.~Yakovlev,
	``QCD sum rules for heavy baryons at next-to-leading order in alpha-s,''
	Phys. Rev. D \textbf{55} (1997), 3016-3026
	[arXiv:hep-ph/9609469 [hep-ph]].
	
	\bibitem{Li:2012nk}
	H.~n.~Li, Y.~L.~Shen and Y.~M.~Wang,
	``Next-to-leading-order corrections to $B \to \pi$ form factors in $k_T$ factorization,''
	Phys. Rev. D \textbf{85}, 074004 (2012)
	[arXiv:1201.5066 [hep-ph]].
	
	
	
	\bibitem{Schlumpf:1992ce}
	F.~Schlumpf, ``Relativistic constituent quark model for baryons,''
	[arXiv:hep-ph/9211255 [hep-ph]].
	
	\bibitem{Hussain:1990uu}
	F.~Hussain, J.~G.~Korner, M.~Kramer and G.~Thompson,
	``On heavy baryon decay form-factors,'' Z. Phys. C \textbf{51}, 321 (1991).
	
	\bibitem{Braun:2000kw}
	V.~Braun, R.~J.~Fries, N.~Mahnke and E.~Stein,
	``Higher twist distribution amplitudes of the nucleon in QCD,''
	Nucl. Phys. B \textbf{589}, 381 (2000) [erratum: Nucl. Phys. B \textbf{607}, 433 (2001)]
	[arXiv:hep-ph/0007279 [hep-ph]].
	
	
	\bibitem{Kurimoto:2001zj}
	T.~Kurimoto, H.~n.~Li and A.~I.~Sanda,
	``Leading power contributions to $B \to \pi, \rho$ transition form-factors,''
	Phys. Rev. D \textbf{65}, 014007 (2002)
	[arXiv:hep-ph/0105003 [hep-ph]].
	
	\bibitem{Li:1992nu}
	H.~n.~Li and G.~F.~Sterman, ``The Perturbative pion form-factor with Sudakov suppression,''
	Nucl. Phys. B \textbf{381}, 129 (1992). 
	
	
	\bibitem{Lu:2002ny}
	C.~D.~Lu and M.~Z.~Yang,
	``B to light meson transition form-factors calculated in perturbative QCD approach,''
	Eur. Phys. J. C \textbf{28}, 515-523  (2003)
	[arXiv:hep-ph/0212373 [hep-ph]].
	
	\bibitem{Mohanta:2000nk}
	R.~Mohanta, A.~K.~Giri and M.~P.~Khanna,
	``Charmless two-body hadronic decays of $\Lambda_b$ baryon,''
	Phys. Rev. D \textbf{63}, 074001 (2001)
	[arXiv:hep-ph/0006109 [hep-ph]].
	
	\bibitem{Khodjamirian:2011jp}
	A.~Khodjamirian, C.~Klein, T.~Mannel and Y.~M.~Wang,
	``Form factors and strong couplings of heavy baryons from QCD light-cone sum rules,''
	JHEP \textbf{09}, 106 (2011)
	[arXiv:1108.2971 [hep-ph]].
	
	\bibitem{Huang:2004vf}
	M.~q.~Huang and D.~W.~Wang,
	``Light cone QCD sum rules for the semileptonic decay $\Lambda_b\to p l \bar\nu$,''
	Phys. Rev. D \textbf{69}, 094003 (2004)
	[arXiv:hep-ph/0401094 [hep-ph]].
	
	\bibitem{Faustov:2016pal}
	R.~N.~Faustov and V.~O.~Galkin,
	``Semileptonic decays of $\Lambda_b$ baryons in the relativistic quark model,''
	Phys. Rev. D \textbf{94}, 073008 (2016)
	[arXiv:1609.00199 [hep-ph]].
	
	\bibitem{Manohar:2000dt}
	A.~V.~Manohar and M.~B.~Wise,
	``Heavy quark physics,''
	Camb. Monogr. Part. Phys. Nucl. Phys. Cosmol. \textbf{10}, 1-191 (2000).
	
	\bibitem{Kindra:2018ayz}
	B.~Kindra and N.~Mahajan,
	``Predictions of angular observables for $\bar{B}_s\to K^{\ast}\ell\ell$ and $\bar{B}\to \rho\ell\ell$ in the standard model,'' 
	Phys. Rev. D \textbf{98}, 094012 (2018) 
	[arXiv:1803.05876 [hep-ph]].
	
	
	
	
	
	
	
	
	
	\bibitem{Aliev:2019thw}
	T.~M.~Aliev, T.~Barakat and M.~Savc\i{},
	``Study of the $\Lambda _{b} \to N^*\ell ^+ \ell ^- $ decay in light cone sum rules,''
	Eur. Phys. J. C \textbf{79}, 383 (2019) 
	[arXiv:1901.04894 [hep-ph]].
	
	
	\bibitem{Azizi:2019tcn}
	K.~Azizi, A.~T.~Olgun and Z.~Tavuko\u{g}lu,
	``Effects of vector leptoquarks on $\Lambda_b \to \Lambda_c\ell\bar \nu_\ell$ decay,''
	Chin. Phys. C \textbf{45}, 013113 (2021)
	[arXiv:1912.03007 [hep-ph]].
	
	
	\bibitem{Azizi:2018axf}
	K.~Azizi and J.~Y.~S\"ung\"u,
	``Semileptonic $\Lambda_{b}\to \Lambda_{c}{\ell}\bar\nu_{\ell}$ Transition in Full QCD,''
	Phys. Rev. D \textbf{97}, 074007 (2018) 
	[arXiv:1803.02085 [hep-ph]].
	
	\bibitem{Zhang:2019jax}
	J.~Zhang, J.~Su and Q.~Zeng, ``Contributions of vector leptoquark to $\Xi_b \to \Xi_c \tau \nu_{\tau}$ decay,'' Nucl. Phys. B \textbf{938}, 131 (2019).
	
	\bibitem{Gutsche:2015mxa}
	T.~Gutsche, M.~A.~Ivanov, J.~G.~K\"orner, V.~E.~Lyubovitskij, P.~Santorelli and N.~Habyl,
	``Semileptonic decay $\Lambda_b \to \Lambda_c + \tau^- + \bar{\nu_\tau}$ in the covariant confined quark model,'' Phys. Rev. D \textbf{91}, 074001 (2015)
	[erratum: Phys. Rev. D \textbf{91}, 119907 (2015)]
	[arXiv:1502.04864 [hep-ph]].
	
	\bibitem{Shivashankara:2015cta}
	S.~Shivashankara, W.~Wu and A.~Datta,
	``$\Lambda_b \to \Lambda_c \tau \bar{\nu}_{\tau}$ Decay in the standard model and with new physics,''
	Phys. Rev. D \textbf{91}, 115003 (2015) 
	[arXiv:1502.07230 [hep-ph]].
	
	\bibitem{Dutta:2015ueb}
	R.~Dutta,
	``$\Lambda_b \to (\Lambda_c,\,p)\,\tau\,\nu$ decays within standard model and beyond,''
	Phys. Rev. D \textbf{93}, 054003 (2016) 
	[arXiv:1512.04034 [hep-ph]].
	
	\bibitem{Zwicky:2013eda}
	R.~Zwicky,
	``Endpoint symmetries of helicity amplitudes,''
	Nucl. Phys. B \textbf{975} (2022), 115673
	[arXiv:1309.7802 [hep-ph]].
	
	\bibitem{Hiller:2021zth}
	G.~Hiller and R.~Zwicky,
	``Endpoint relations for baryons,''
	JHEP \textbf{11} (2021), 073
	[arXiv:2107.12993 [hep-ph]].
	
	\bibitem{LHCb:2015eia}
	R.~Aaij \textit{et al.} [LHCb Collab.],
	``Determination of the quark coupling strength $|V_{ub}|$ using baryonic decays,''
	Nature Phys. \textbf{11}, 743 (2015) 
	[arXiv:1504.01568 [hep-ex]].
	
	\bibitem{ParticleDataGroup:2020ssz}
	P.~A.~Zyla \textit{et al.} [Particle Data Group], ``Review of Particle Physics,''
	PTEP \textbf{2020}, 083C01 (2020). 
	
	
	
\end{thebibliography}
\end{document}